\documentclass[12pt,letterpaper]{article}
\pdfoutput =1
\usepackage[pdftex]{graphicx,color}
\usepackage{hyperref}
\input{psfig}
\usepackage{amsmath,amssymb}
\usepackage[dvips,letterpaper,text={6.5in,9in}]{geometry}
\usepackage{fancyhdr}
\usepackage{verbatim}

\newcommand\ltap{\
  \raise.3ex\hbox{$<$\kern-.75em\lower1ex\hbox{$\sim$}}\ }
\newcommand\gtap{\
  \raise.3ex\hbox{$>$\kern-.75em\lower1ex\hbox{$\sim$}}\ }

\newcommand\simge{\mathrel{%
   \rlap{\raise 0.511ex \hbox{$>$}}{\lower 0.511ex \hbox{$\sim$}}}}
\newcommand\simle{\mathrel{
   \rlap{\raise 0.511ex \hbox{$<$}}{\lower 0.511ex \hbox{$\sim$}}}}

\newcommand{\slashchar}[1]%
        {\kern .25em\raise.18ex\hbox{$/$}\kern-.75em #1}
\def\lsim{\mathrel{\raise.3ex\hbox{$<$\kern-.75em\lower1ex\hbox{$\sim$}}}}
\def\gsim{\mathrel{\raise.3ex\hbox{$>$\kern-.75em\lower1ex\hbox{$\sim$}}}}
\newcommand{\bs}{\boldsymbol}

\newcommand\CL{{\cal L}}

\newcommand\CO{{\cal O}}

\newcommand\be{\begin{equation}}
\newcommand\ee{\end{equation}}
\newcommand\bea{\begin{eqnarray}}
\newcommand\eea{\end{eqnarray}}
\newcommand\ba{\begin{array}}
\newcommand\ea{\end{array}}
\newcommand\nn{\nonumber}

\newcommand{\thalf}{\textstyle{\frac{1}{2}}}

\newcommand{\tthird}{\textstyle{\frac{1}{3}}}

\newcommand{\tfourth}{\textstyle{\frac{1}{4}}}

\newcommand\ra{\rightarrow}

\newcommand\gev{{\rm GeV}}
\newcommand\tev{{\rm TeV}}

\newcommand\pb{{\rm pb}}

\newcommand\fb{{\rm fb}}
\newcommand\ifb{{\rm fb}^{-1}}

\newcommand\etmiss{\slashchar{E}_T}

\newcommand\ellm{\ell^-}
\newcommand\ellpm{\ell^\pm}
\newcommand\ellp{\ell^+}

\newcommand\atc{\alpha_{TC}}

\newcommand\Ltc{\Lambda_{TC}}

\newcommand\grpp{g_{\rho_T\pi_T\pi_T}}

\newcommand\tom{\omega_{T}}
\newcommand\tro{\rho_{T}}
\newcommand\atro{\alpha_{\rho_T}}

\newcommand\ta{a_T}

\newcommand\tapm{a_T^\pm}

\newcommand\tropm{\rho_{T}^\pm}

\newcommand\tpi{\pi_T}
\newcommand\tpipm{\pi_T^\pm}

\newcommand\tpiz{\pi_T^0}

\newcommand\jets{{\rm jets}}

\newcommand\Mjj{M_{jj}}

\newcommand\MWjj{M_{Wjj}}
\newcommand\MZjj{M_{Zjj}}
\newcommand\ptjj{p_{T}(jj)}

\newcommand\dRll{\Delta R_{\ell\ell}}
\newcommand\dXll{\Delta\chi_{\ell\ell}}

\newcommand\dphi{\Delta\phi}
\newcommand\deta{\Delta\eta}
\newcommand\dR{\Delta R}
\newcommand\dX{\Delta\chi}
\newcommand\dRm{(\Delta R)_{\rm min}}
\newcommand\dXm{(\Delta\chi)_{\rm min}}

\newcommand\cth{c_{\theta}}
\newcommand\sth{s_{\theta}}
\newcommand\cthst{c_{\theta^*}}
\newcommand\sthst{s_{\theta^*}}
\newcommand\cphst{c_{\phi^*}}
\newcommand\sphst{s_{\phi^*}}
\newcommand\bth{b_{\theta}}
\newcommand\bthst{b_{\theta^*}}
\newcommand\bphst{b_{\phi^*}}

\begin{document}

\title{
\vskip -15mm
\begin{flushright}
 \vskip -15mm
 {\small FERMILAB-PUB-12-015-T\\
   LAPTh-003/12\\
 }
 \vskip 5mm
 \end{flushright}
{\Large{\bf Testing the Technicolor Interpretation \\of CDF's Dijet
    Excess at the LHC}}\\
} \author{
  {\large Estia Eichten$^{1}$\thanks{eichten@fnal.gov} ,\,
  Kenneth Lane$^{2}$\thanks{lane@physics.bu.edu},\,
  Adam Martin$^{1}$\thanks{aomartin@fnal.gov}}\, and 
  Eric Pilon$^{3}$\thanks{pilon@lapp.in2p3.fr}\\
{\large {$^{1}$}Theoretical Physics Group, Fermi National Accelerator
  Laboratory}\\
{\large P.O. Box 500, Batavia, Illinois 60510}\\
{\large $^{2}$Department of Physics, Boston University}\\
{\large 590 Commonwealth Avenue, Boston, Massachusetts 02215}\\
{\large $^{3}$Laboratoire d'Annecy-le-Vieux de Physique Th\'eorique} \\
{\large UMR5108\,, Universit\'e de Savoie, CNRS} \\
{\large B.P.~110, F-74941, Annecy-le-Vieux Cedex, France}\\
}
\maketitle

\begin{abstract}
  
  Under the assumption that the dijet excess seen by the CDF Collaboration
  near $150\,\gev$ in $Wjj$ production is due to the lightest technipion of
  the low-scale technicolor process $\tro \ra W\tpi$, we study its
  observability in LHC detectors with $1$--$20\,\ifb$ of data. We describe
  interesting new kinematic tests that can provide independent confirmation
  of this LSTC hypothesis. We find that cuts similar to those employed by
  CDF, and recently by ATLAS, cannot confirm the dijet signal. We propose
  cuts tailored to the LSTC hypothesis and its backgrounds at the LHC that
  may reveal $\tro \ra \ell\nu jj$. Observation of the isospin-related
  channel $\tropm \ra Z\tpipm \ra \ellp\ellm jj$ and of $\tropm \ra WZ$ in
  the $\ellp\ellm\ellpm\nu_\ell$ and $\ellp\ellm jj$ modes will be important
  confirmations of the LSTC interpretation of the CDF signal. The $Z\tpi$
  channel is experimentally cleaner than $W\tpi$ and its rate is known from
  $W\tpi$ by phase space. It can be discovered or excluded with the collider
  data expected in 2012. The $WZ \ra 3\ell\nu$ channel is cleanest of all and
  its rate is determined from $W\tpi$ and the LSTC parameter $\sin\chi$. This
  channel and $WZ \to \ellp\ellm jj$ are discussed as a function of
  $\sin\chi$.

\end{abstract}


\newpage

\section*{1. Introduction}

The CDF Collaboration has reported evidence for a resonance near $150\,\gev$
in the dijet-mass spectrum, $\Mjj$, of $Wjj$ production. This is based on an
integrated luminosity of $4.3\,\ifb$~\cite{Aaltonen:2011mk} and updated with
a total data sample of $7.3\,\ifb$~\cite{CDFnew}. In Ref.~~\cite{CDFnew}, the
resonant dijet excess has a significance of $4.1\,\sigma$.  The D\O\
Collaboration, on the other hand, published a search for this resonance based
on $4.3\,\ifb$ that found no significant excess. Based on a $W+$Higgs boson
production model, D\O\ reported a cross section for a potential signal of
$0.82^{+ 0.83}_{- 0.82}\,\pb$ and a 95\% confidence level upper limit of
$1.9\,\pb$~\cite{Abazov:2011af}. Analyzing its data with the same production
model, CDF reported a signal rate of $3.0\pm 0.7\,\pb$ and a discrepancy
between the two experiments of $2.5\,\sigma$~\cite{AnnoviLP11}. This
discrepancy remains. The purpose of this paper is to urge that the LHC
experiments, ATLAS and CMS, mount searches to test for the CDF dijet excess
in the $Wjj$ and closely related channels. We do this in the context
low-scale technicolor (LSTC), interpreting CDF's dijet excess as the lightest
technipion $\pi_T^{\pm,0}$ of this scenario, produced in association with
$W^\pm$ in the decay $\rho_T^{\pm,0} \ra W\tpi$~\cite{Eichten:2011sh}. The
related channels supporting this interpretation are $\tropm \ra Z\tpipm$ and
$W^\pm Z$. They require the fewest additional LSTC model assumptions to
determine LHC production rates. We assume $\sqrt{s} = 7\,\tev$ and consider
$\int\CL dt = 1$--$20\,\ifb$, an amount of data expected in 2012.\footnote{A
  preliminary version of this paper was circulated as
  Ref.~\cite{Eichten:2011xd}, assuming $\int\CL dt = 1$--$5\,\ifb$. The
  simulations we present here for various fixed luminosities may be applied
  to different ones by scaling the event rates. The question of raising the
  LHC collision energy to $8\,\tev$ in 2012 is under consideration. This
  probably will not be advantageous for the LSTC signals discussed here and
  elsewhere because backgrounds, especially gluon-induced ones, grow faster
  with energy than do the signals.}

Low-scale technicolor (LSTC) is a phenomenology based on walking
technicolor~\cite{Holdom:1981rm, Appelquist:1986an,Yamawaki:1986zg,
  Akiba:1986rr}. The gauge coupling $\atc$ must run very slowly for 100s of
TeV above the TC scale $\Ltc \sim$ several $100\,\gev$ so that extended
technicolor (ETC) can generate sizable quark and lepton masses while
suppressing flavor-changing neutral current
interactions~\cite{Eichten:1979ah}. This may be achieved, e.g., with
technifermions belonging to higher-dimensional representations of the TC
gauge group. The constraints of Ref.~\cite{Eichten:1979ah} on the number of
ETC-fermion representations then imply that there will be technifermions in
the fundamental TC representation as well. They are expected to condense at
an appreciably lower energy scale than those belonging to the
higher-dimensional representations and, thus, their technipions' decay
constant $F_1^2 \ll F_\pi^2 = (246\,\gev)^2$~\cite{Lane:1989ej}. Meson bound
states of these technifermions will have a quarkonium-like spectrum with
masses well below a TeV --- greater than the previous Tevatron limit
$M_{\tro} \simge 250\,\gev$~\cite{Abazov:2006iq, Aaltonen:2009jb} and
probably less than 600--700~GeV, a scale at which we believe the notion of
``low-scale'' TC ceases to make sense. The most accessible states are the
lightest technivectors, $V_T = \tro(I^G J^{PC} = 1^+1^{--})$,
$\tom(0^-1^{--})$ and $\ta(1^-1^{++})$. Through their mixing with the
electroweak bosons, they are readily produced as $s$-channel resonances via
the Drell-Yan process in colliders. Technipions $\tpi(1^-0^{-+})$ are
accessed in $V_T$ decays. A central assumption of LSTC is that these lightest
technihadrons may be treated in isolation, without significant mixing or
other interference from higher-mass states.  Also, we expect that (1) the
lightest technifermions are $SU(3)$-color singlets, (2) isospin violation is
small for $V_T$ and $\tpi$, (3) $M_{\tom} \cong M_{\tro}$, and (4) $M_{\ta}$
is not far above $M_{\tro}$. This last assumption is made to keep the
low-scale TC contribution to the $S$-parameter small. An extensive discussion
of LSTC, including these points and precision electroweak constraints, is
given in Ref.~\cite{Lane:2009ct}.

Walking technicolor has another important consequence: it enhances $M_{\tpi}$
relative to $M_{\tro}$ so that the all-$\tpi$ decay channels of the $V_T$ are
likely to be closed~\cite{Lane:1989ej}. Principal $V_T$-decay modes are
$W\tpi$, $Z\tpi$, $\gamma \tpi$, a pair of EW bosons (which can include one
photon), and fermion-antifermion pairs~\cite{Lane:2002sm,
  Eichten:2007sx,Lane:2009ct}. If allowed by isospin, parity and angular
momentum, $V_T$ decays to one or more weak bosons involve
longitudinally-polarized $W_L/Z_L$, the technipions absorbed via the Higgs
mechanism. These nominally strong decays are suppressed by powers of
$\sin\chi = F_1/F_\pi \ll 1$. This is an important parameter in LSTC. It is a
mixing factor that measures the amount that the lowest-scale technipion is the
mass eigenstate $\tpi$ ($\cos\chi$) and the amount that it is $W_L/Z_L$
($\sin\chi$). Thus, each replacement of a mass-eigenstate $\tpi$ by $W_L/Z_L$
in a $V_T$ decay amplitude costs a factor of $\tan\chi$. Decays to
transversely-polarized $\gamma,W_\perp,Z_\perp$ are suppressed by $g,g'$.
Thus, the $V_T$ are {\em very} narrow, $\Gamma(\tro) \simle 1\,\gev$ and
$\Gamma(\tom,\ta) \simle 0.1\,\gev$ for the masses considered here. These
decays have striking signatures, visible above backgrounds within a limited
mass range at the Tevatron and probably up to 600--700~GeV at the
LHC~\cite{Brooijmans:2008se, Brooijmans:2010tn}.

In Ref.~\cite{Eichten:2011sh} we proposed that CDF's dijet excess is due to
resonant production of $W\tpi$ with $M_{\tpi} = 160\,\gev$.  We took
$M_{\tro} = 290\,\gev$ and $M_{\ta} = 1.1 M_{\tro} = 320\,\gev$. Then, about
75\% of the $W\tpi$ rate at the Tevatron is due to $\tro \ra W\tpi$ and, of
this, most of the $W$'s are longitudinally polarized.\footnote{About 70\% of
  the $W\tpi$ rate at the LHC is due to the $\tro$.} The remainder is
dominated by $\ta$ production. Its decay, and a small fraction of the
$\tro$'s, involve $W_\perp$ production, which is generated by dimension-five
operators~\cite{Lane:2009ct}. These operators are suppressed by mass
parameters $M_{V,A}$ that we take equal to $M_{\tro}$. The other LSTC
parameters relevant to $W\tpi$ production are $\grpp$ and $\sin\chi$. The
$\tro \to \tpi\tpi$ coupling $\grpp$ is the same for all $\tro$ decays
considered here and it is simply scaled from QCD; its {\sc Pythia} default
value is $\atro = \grpp^2/4\pi = 2.16(3/N_{TC})$ with $N_{TC} = 4$. We use
$\sin\chi = 1/3$. Using the LSTC model implemented in {\sc
  Pythia}~\cite{Lane:2002sm,Eichten:2007sx, Sjostrand:2006za}, we found
$\sigma(\bar pp \ra \tro \ra W\tpi \ra Wjj) = 2.2\,\pb$ ($480\,\fb$ after $W
\ra e\nu,\,\mu\nu$).\footnote{This includes $B(\tpi \ra \bar q q) \simeq
  90\%$ in the default {\sc Pythia} $\tpi$-decay table.} Using CDF's cuts, we
closely matched its $\Mjj$ distribution for signal and background. Motivated
by the peculiar kinematics of $\tro$ production at the Tevatron and $\tro \ra
W\tpi$ decay, we also suggested cuts intended to enhance the $\tpi$ signal's
significance and to make $\tro \ra Wjj$ visible. Several distributions of
data in the excess region $115\,\gev < \Mjj < 175\,\gev$ published by
CDF~\cite{CDFnew} --- notably $\MWjj$, $\ptjj$, $\dphi$ and $\dR =
\sqrt{(\deta)^2 + (\dphi)^2}$ --- fit the expectations of the LSTC model very
well. The background-subtracted $\dR$ distribution, in particular, has a
behavior which, we believe, furnishes strong support for our dijet production
mechanism.

The purpose of this paper is to propose and study ways to test for the CDF
signal at the LHC. In Sec.~2 we review the kinematics of $\tro, \ta \ra
W\tpi$ and $Z\tpi$ in LSTC. We also present an interesting new result: the
nonanalytic behavior of $d\sigma/d(\dR$) and $d\sigma/d(\dX)$ at their
thresholds, $\dRm$ and $\dXm$. Here $\dX$ is the opening angle between the
$\tpi$ decay jets in the $\tro$ rest frame. For massless jets, a good
approximation, we find that $\dRm = \dXm = 2\cos^{-1}(v)$, where $v =
p_{\tpi}/E_{\tpi}$ is the $\tpi$ velocity in the $\tro$ rest frame. This
result, peculiar to production models such as LSTC in which a narrow
resonance decays to another narrow resonance plus a $W$ or $Z$, provides
measures of $v$ independent of $p/E$ and, hence, valuable corroboration of
this type of production. In Sec.~3 we consider the $\tro,\ta \ra W\tpi$
process. Its LHC cross section is $7.9\,\pb$ but, for CDF cuts, its
backgrounds have increased by about a factor of ten. This makes testing for
the dijet excess in this channel very challenging. We suggest cuts which
enhance signal-to-background $(S/B)$ but which will still require a very good
understanding of the backgrounds in $Wjj$ production and perhaps luminosity
$\simge 10\,\ifb$ to observe, or exclude, this signal. In Sec.~4 we study
$\tropm,\tapm \ra Z\tpipm$, whose cross section is $2.3\,\pb$ at $7\,\tev$
($155\,\fb$ after $Z \ra e^+e^-,\, \mu^+\mu^-$). This is the isospin partner
of $\tropm,\tapm \ra W\tpiz$, so its cross section is rather confidently
known. The $\ellp\ellm jj$ channel is free of QCD multijet and $\bar t t$
backgrounds and missing energy uncertainty. Reconstructing the $Zjj$
invariant mass and other signal distributions, particularly in $\dR$ and
$\dX$, will benefit from this. {\em Because of these features, we believe
  that the $Z\tpi \ra Zjj$ mode will be the surest test of CDF's dijet signal
  at the LHC.} In Sec.~5, we study $\tropm,\tapm \ra WZ$. The cross section
for this mode is proportional to $\tan^2\chi$ times the $\tropm,\tapm \ra
W^\pm\tpiz$ and $Z\tpipm$ rates, but enhanced by its greater phase space. We
predict $\sigma(\tropm,\tapm \ra WZ) = 1.3\,\pb$ for $\sin\chi = 1/3$. In the
all-leptons mode (with $e$'s and $\mu$'s), the rate is only $20\,\fb$, but
jet-related uncertainties are entirely absent. The $WZ \to \ellp\ellm jj$
mode is also an interesting target of opportunity so long as $\sin\chi
\simge 1/4$.  The $\dR$ and $\dX$ distributions for $Z \to jj$ again provide
support for our narrow LSTC-resonance production model. In short, one or both
of the $Z\tpi$ and $WZ$ modes should be accessible with the $\simge 20\,\ifb$
expected to be in hand by the end of 2012.  Finally, we present in an
appendix the details of calculations in Sec.~2 regarding the nonanalytic
threshold behavior of the $\dX$ and $\dR$ distributions.

While the simulations of the CDF signal in this paper are made in the context
of low-scale technicolor, their qualitative features apply to any model in
which that signal is due to $\bar q q$ production of a narrow resonance
decaying to a $W$ plus another narrow resonance. Several papers have appeared
proposing such an $s$-channel mechanism~\cite{Kilic:2011sr,Cao:2011yt,
  Chen:2011wp, Fan:2011vw, Ghosh:2011np,Gunion:2011bx}. With similar
resonance masses to our LSTC proposal, these models will have kinematic
distributions like those we describe in Sec.~2. However, not all these models
will have the $Zjj$ and $WZ$ signals of LSTC. There are also a large number
of papers proposing that the CDF signal is due to production of a new
particle (e.g., a leptophobic $Z'$) that is not resonantly
produced~\cite{Buckley:2011vc, Hewett:2011fk, Harnik:2011mv, Dobrescu:2011fk,
  Nelson:2011us, Yu:2011cw, Cheung:2011zt}. These ``$t$-channel'' models will
not pass our kinematic tests.

\section*{2. LSTC Kinematics and Threshold Nonanalyticity}

The kinematics of $\tro \ra W\tpi$ at the Tevatron and LHC are a consequence
of the basic LSTC feature that walking TC enhancements of $M_{\tpi}$ strongly
suggest $M_{\tro} < 2M_{\tpi}$ and, indeed, that the phase space for $\tro
\ra W\tpi$ is quite limited~\cite{Lane:1989ej, Eichten:1997yq}. At the
Tevatron, a $290\,\gev$ $\tro$ is produced almost at rest, with almost no
$p_T$ and very little boost along the beam direction. At the LHC, $p_T(\tro)
\simle 25\,\gev$ and $\eta(\tro) \simle 2.0$. Furthermore, the $\tpi$ is
emitted very slowly in the $\tro$ rest frame --- $v \simeq 0.4$ for our
assumed masses --- so that its decay jets are roughly back-to-back in the lab
frame. Thus, $p_T(\tpi) \simle 80\,\gev$ and the $z$-boost invariant
quantities $\dphi$ and $\dR = \sqrt{(\deta)^2 + (\dphi)^2}$ are peaked at
large values less than $\pi$.

These features of LSTC are supported by CDF's $7.3\,\ifb$
data~\cite{CDFnew}. Figures~\ref{fig:CDFMWjj}--\ref{fig:CDFdRjj} show
distributions before and after background subtraction taken from the $115 <
\Mjj < 175\,\gev$ region containing the dijet excess. The subtracted-data
$\MWjj$ signal has a narrow resonant shape quite near
$290\,\gev$. Unfortunately, the background peaks just below that mass so that
one may be concerned that the subtracted data's peak is due to
underestimating the background. Also, as we expect, the subtracted $p_T(jj)$
data falls off sharply above $75\,\gev$ and the subtracted $\dphi$ data is
strongly peaked at large values. Again, one may worry that these are
artifacts of the peaks of the $\MWjj$ background and the position of the
$\Mjj$ excess.

\begin{figure}[!t]
 \begin{center}
\includegraphics[width=3.15in, height=3.15in]{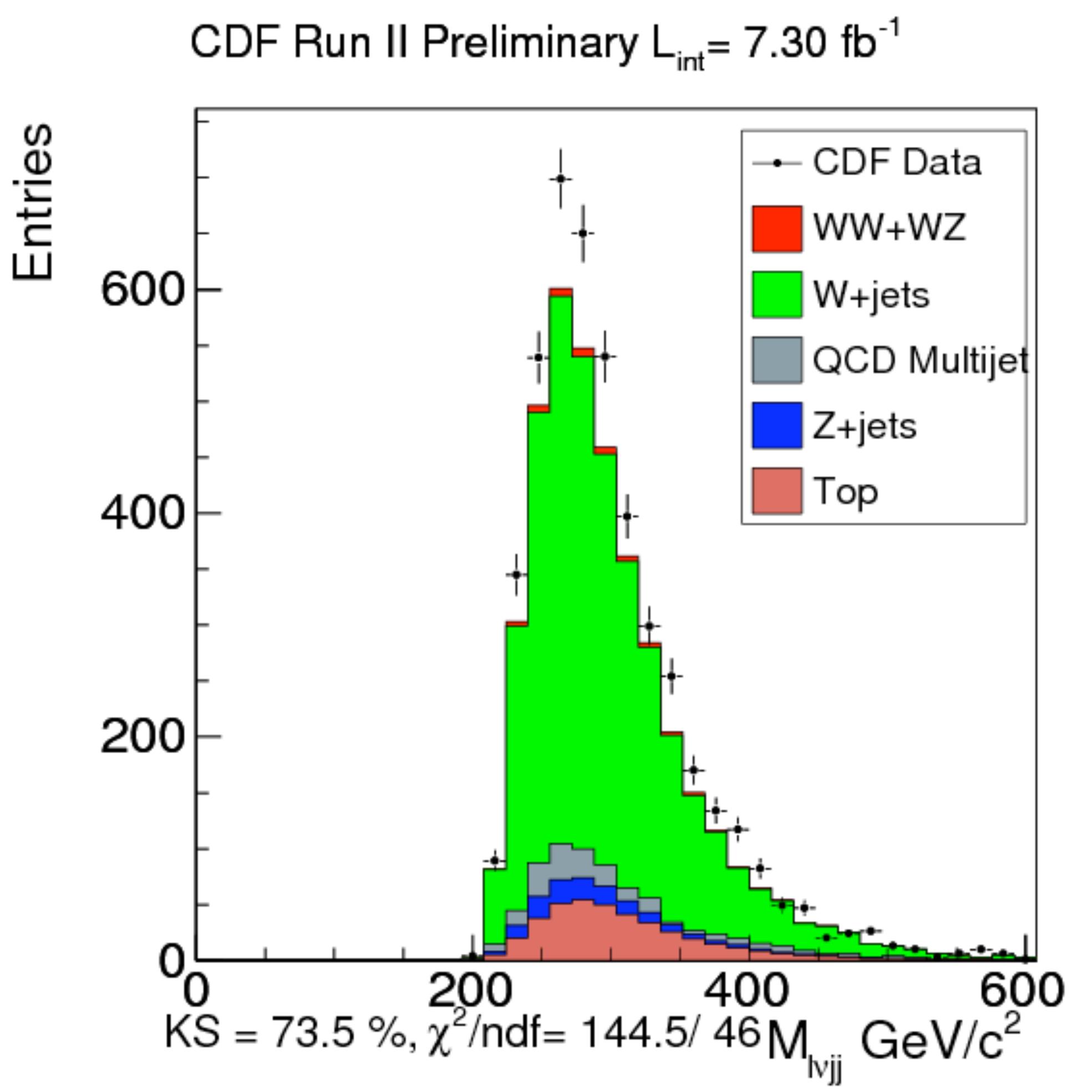}
\includegraphics[width=3.15in, height=3.15in]{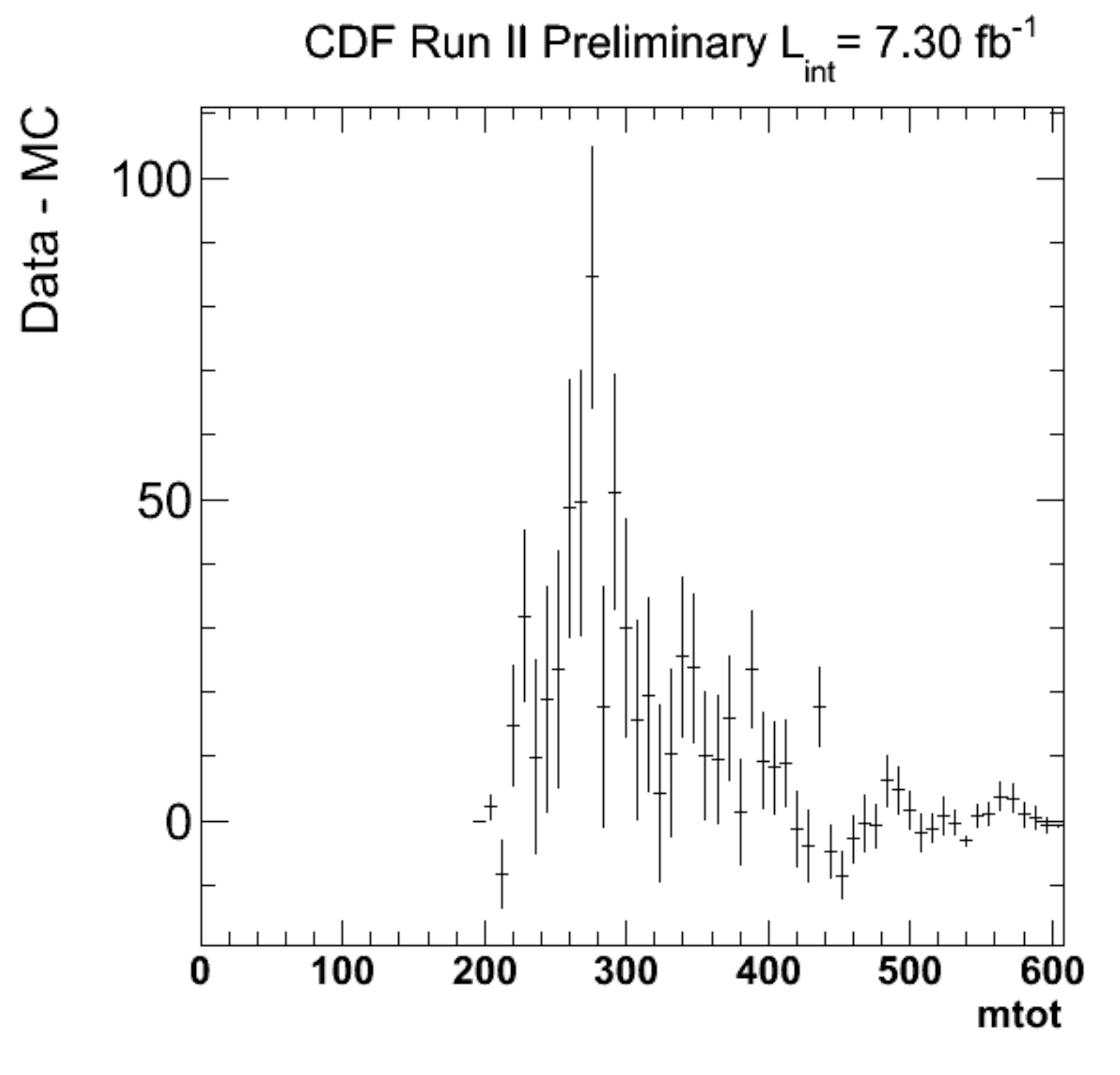}
\caption{CDF $\MWjj$ distributions for $\int \CL dt =
  7.3\,\ifb$ from the dijet signal region $115 < \Mjj <
  175\,\gev$~\cite{CDFnew}. Left: Expected backgrounds and data; right:
  background subtracted data.}
  \label{fig:CDFMWjj}
 \end{center}
 \end{figure}
\begin{figure}[!ht]
 \begin{center}
\includegraphics[width=3.15in, height=3.15in]{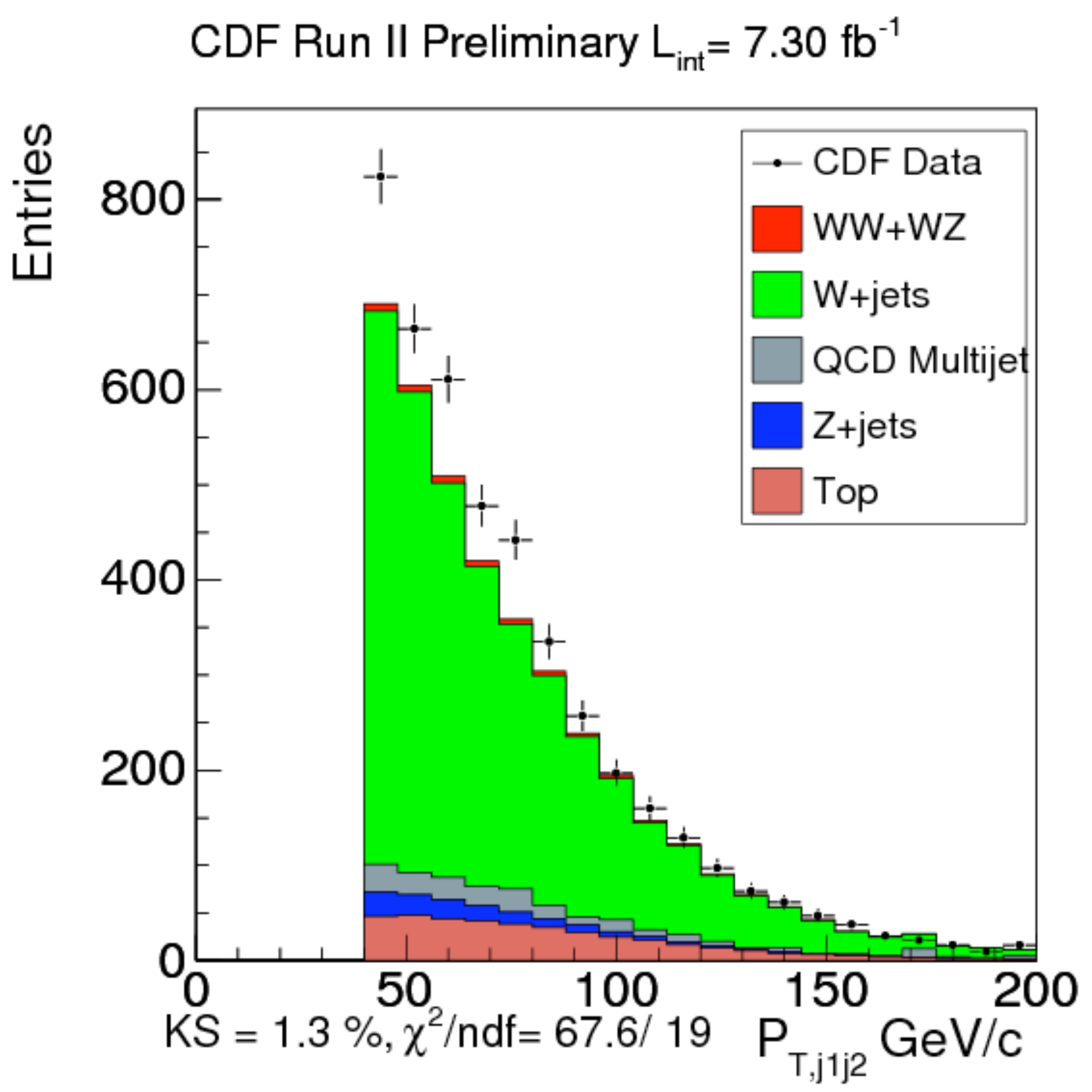}
\includegraphics[width=3.15in, height=3.15in]{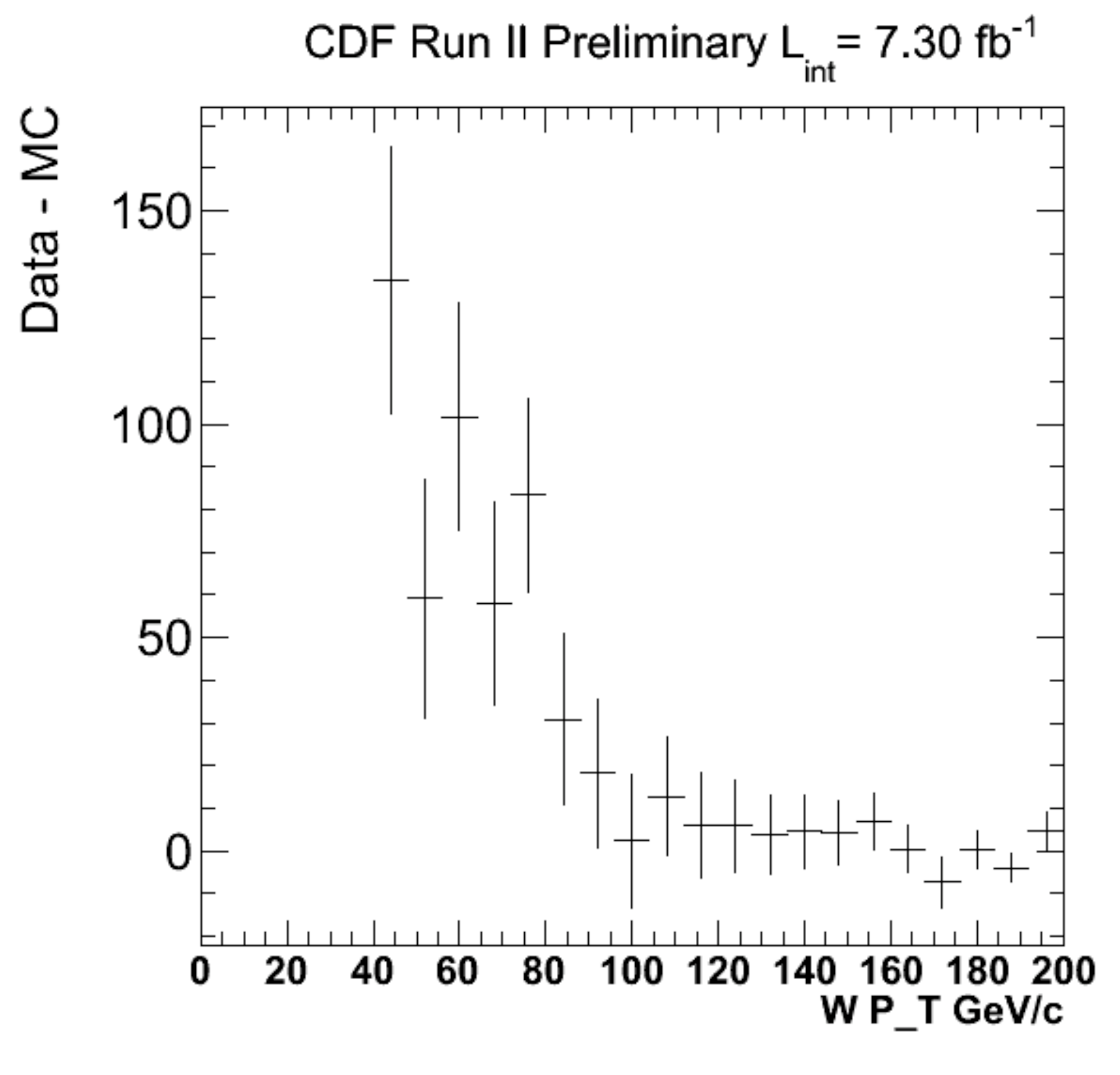}
\caption{CDF $p_T(jj)$ distributions for $\int \CL dt =
  7.3\,\ifb$ from the dijet signal region $115 < \Mjj <
  175\,\gev$~\cite{CDFnew}. Left: Expected backgrounds and data; right:
  background subtracted data.}
  \label{fig:CDFpTjj}
 \end{center}
 \end{figure}

 The background-subtracted $\dR$ distribution, however, is very interesting.
 It is practically zero for $\dR < 2.25$, then rises sharply to a broad
 maximum before falling to zero again at $\dR \simeq 3.5$. This
 behavior, and a somewhat similar one we predict for $\dX$ are the main
 subject of this section. We will show that the threshold form of the $\dR$
 and $\dX$ distributions provide direct measures of the velocity of the dijet
 system in the subprocess center-of-mass frame that are independent of
 measuring $p/E$ and, thus, are independent checks on the two-resonance
 topology of the dijet's production mechanism.\footnote{Recall that $\dX$ is
   defined in the $\tro$ rest frame, while $\dR$ is defined in the lab frame.
   If one wishes to remove the effect of $p_T(\tro)$ on $\dR$, it should be
   defined in the $\tro$ frame.} One might think that the corresponding
 $\dRll$ and $\dXll$ distributions from $Z \to \ellp\ellm$ would be similarly
 valuable. Unfortunately, because the dileptons come from real $Z$'s and our
 cuts make the background $Z$'s like the signal ones, $\dRll$ and $\dXll$ are
 indistinguishable from their backgrounds.

\begin{figure}[!t]
 \begin{center}
\includegraphics[width=3.15in, height=3.15in]{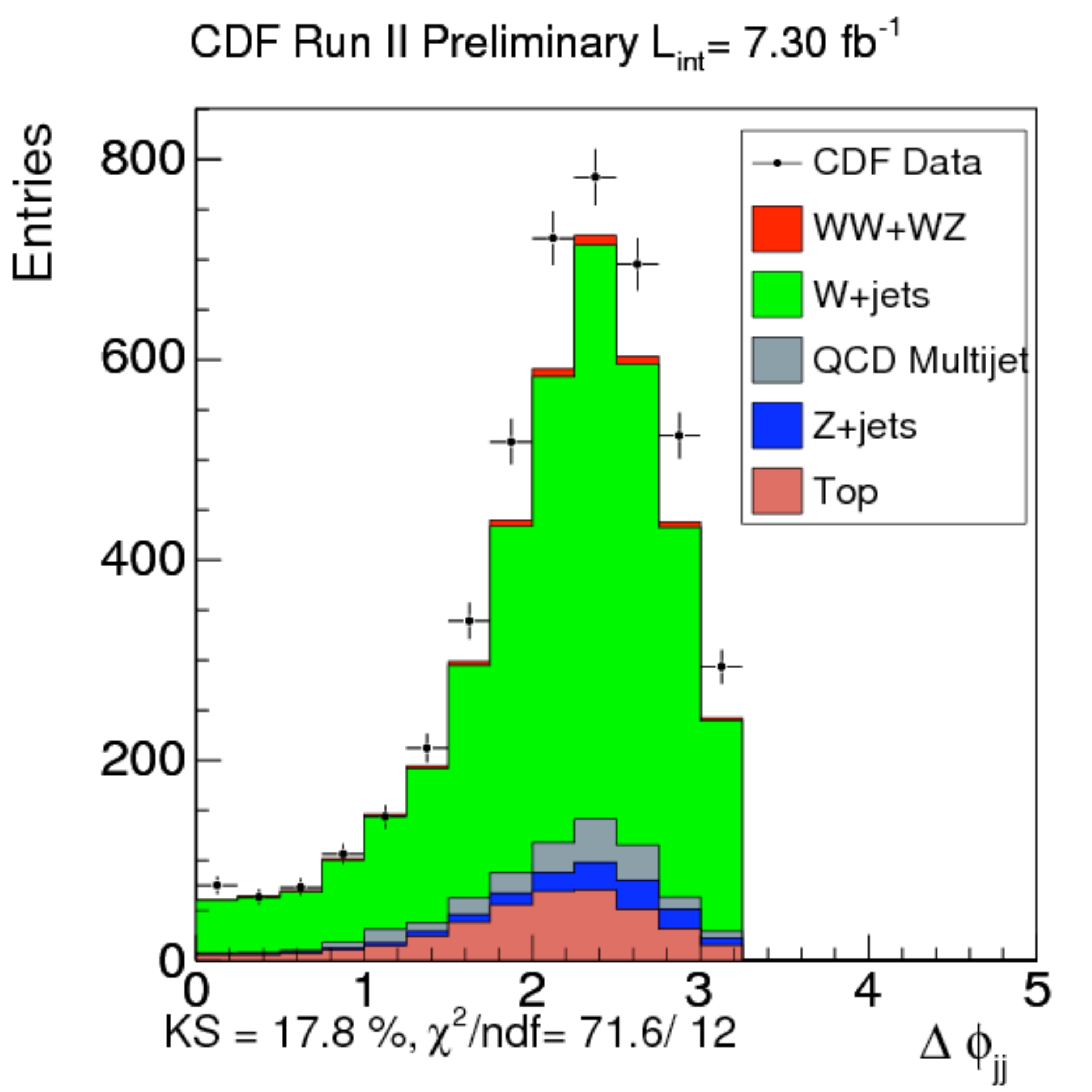}
\includegraphics[width=3.15in, height=3.15in]{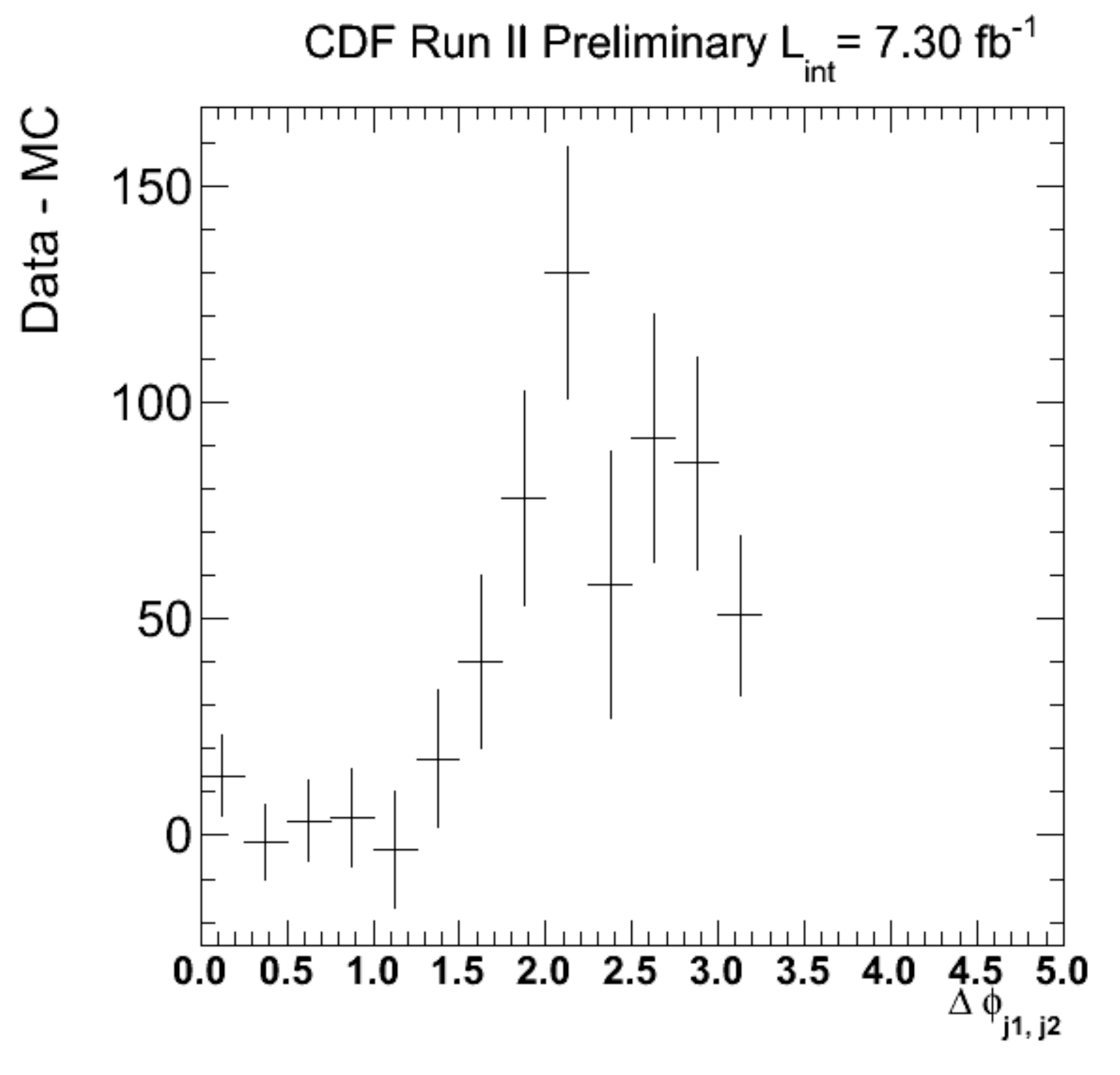}
\caption{CDF $\dphi$ distributions for $\int \CL dt =
  7.3\,\ifb$ from the dijet signal region $115 < \Mjj <
  175\,\gev$~\cite{CDFnew}. Left: Expected backgrounds and data; right:
  background subtracted data.}
  \label{fig:CDFdphijj}
 \end{center}
 \end{figure}
\begin{figure}[!ht]
 \begin{center}
\includegraphics[width=3.15in, height=3.15in]{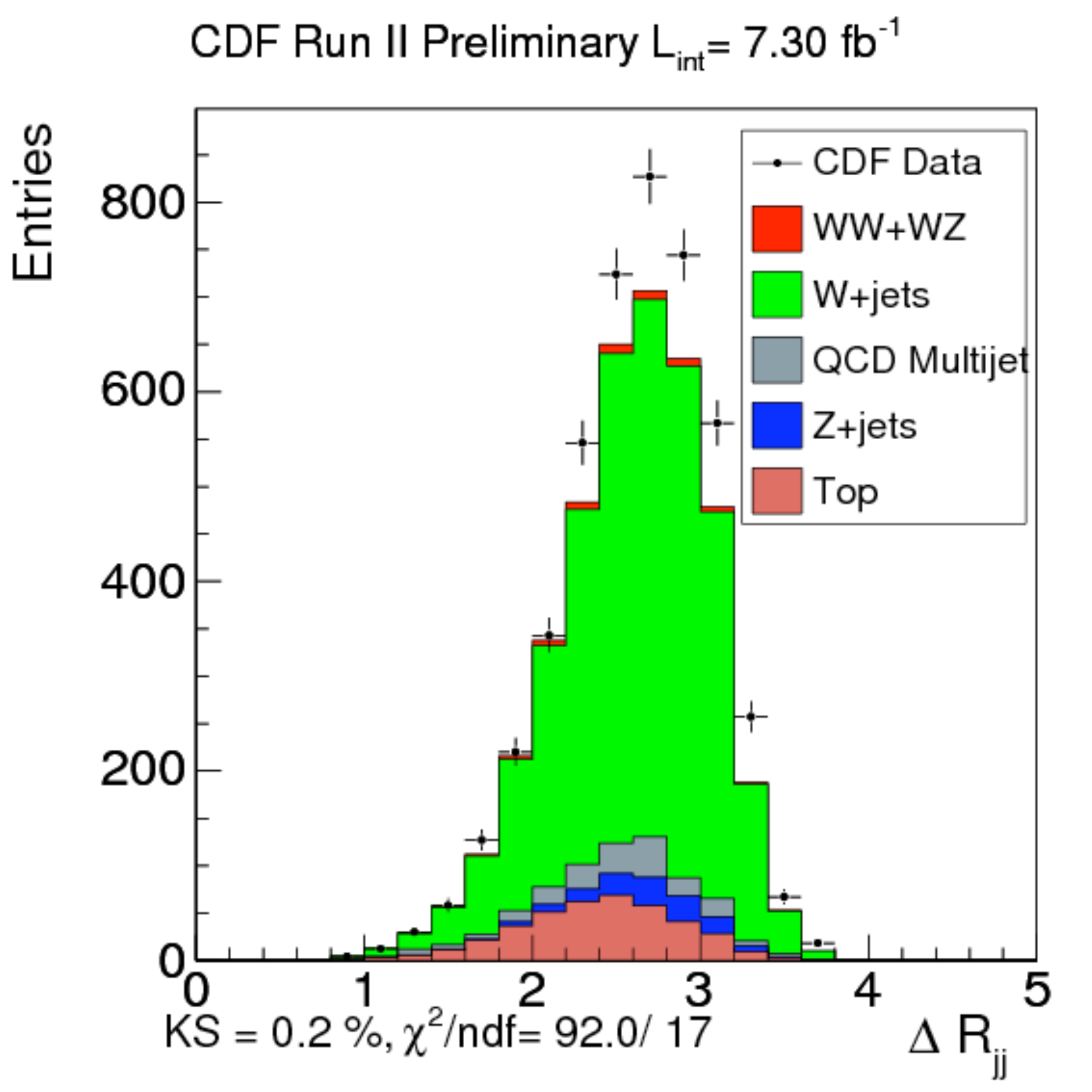}
\includegraphics[width=3.15in, height=3.15in]{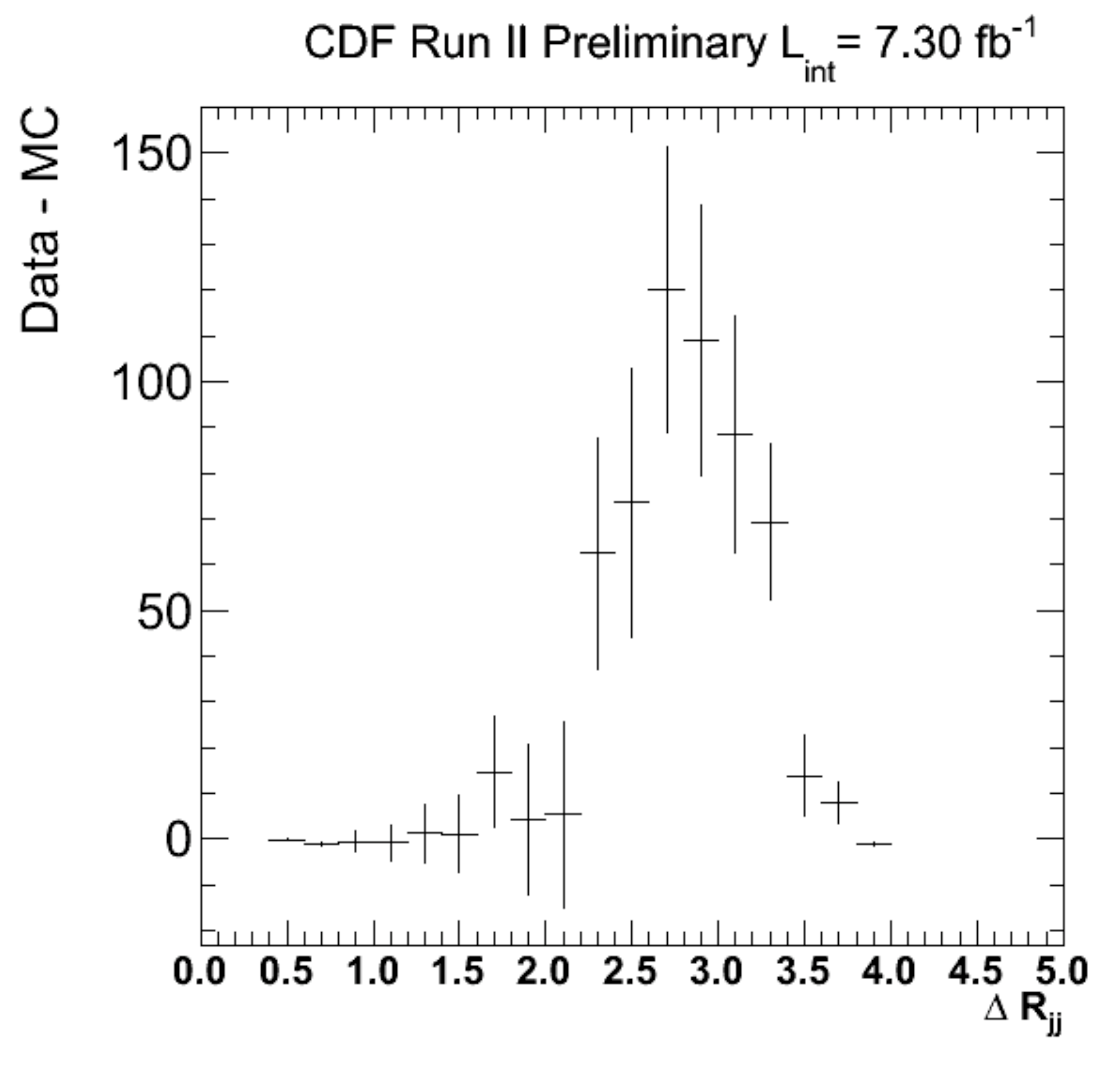}
\caption{CDF $\dR$ distributions for $\int \CL dt =
  7.3\,\ifb$ from the dijet signal region $115 < \Mjj <
  175\,\gev$~\cite{CDFnew}. Left: Expected backgrounds and data; right:
  background subtracted data.}
  \label{fig:CDFdRjj}
 \end{center}
 \end{figure}

 For our analysis, we assume the jets from $\tpi$ decay are massless. We have
 examined the effect of including jet masses and found them to be
 unimportant. We will remark briefly on them at the end of this section. We
 first consider the dominant $\tro$ contribution to $W/Z\tpi$ production,
 commenting on the $\ta$ contribution also at the end.
 
 Define the angles $\theta$, $\theta^*$ and $\phi^*$ as follows: Choose the
 $z$-axis as the direction of the event's boost; this is usually the direction
 of the incoming quark in the subprocess c.m.~frame. In the $\tro$ rest
 frame, $\theta$ is the polar angle of the $\tpi$ velocity ${\bs v}$, the
 angle it makes with the $z$-axis.  Define the $xz$-plane as the one
 containing the unit vectors $\hat{\bs z}$ and $\hat{\bs v}$, so that
 $\hat{\bs v} = \hat{\bs x}\sin\theta + \hat{\bs z}\cos\theta$, and $\hat{\bs
   y} = \hat{\bs z} \times \hat{\bs x}$. Define a starred coordinate system
 {\em in the $\tpi$ rest frame} by making a rotation by angle~$\theta$ about
 the $y$-axis of the $\tro$ frame. This rotation takes $\hat{\bs z}$ into
 $\hat{\bs z}^* = \hat{\bs v}$ and $\hat{\bs x}$ into $\hat{\bs x}^* =
 \hat{\bs x}\cos\theta - \hat{\bs z}\sin\theta$. In this frame, let $\hat
 {\bs p}_1^*$ be the unit vector in the direction of the jet (parton) making
 the smaller angle with the direction of $\hat{\bs v}$.  This angle is
 $\theta^*$; the azimuthal angle of ${\bs p}_1^* = -{\bs p}_2^*$ is $\phi^*$:
\be\label{eq:angles}
\cos\theta = \hat{\bs z}\cdot \hat{\bs v}, \quad
\cos\theta^* = \hat{\bs p}_1^*\cdot \hat{\bs v}, \quad
\tan\phi^* = p_{1y^*}^*/p_{1x^*}^*.
\ee
Note that, since $\tpi \ra \bar q q$ is isotropic in its
rest frame, the distribution $d\sigma(\bar q q \ra \tro \ra
Wjj)/d(\cos\theta^*) = \sigma/2$, where $\sigma$ is the total subprocess
cross section.

It is easier to consider the $d\sigma/d(\dX)$ distribution first. For
massless jets,
\be\label{eq:cosdchi}
1-\cos(\dX) = \frac{2(1-v^2)}{1-v^2\cos^2\theta^*}\,.
\ee
The minimum value of $\dX$ occurs when $\theta^* = \pi/2$ (i.e., ${\bs v}
\perp {\bs p}_1^*$), and so
\be\label{eq:delchimin}
\pi \ge \dX \ge \dXm = 2\cos^{-1}(v)\,.
\ee
From Eq.~(\ref{eq:cosdchi}), it is easy to see that
\be
\frac{d\sigma}{d(\dX)} = \frac{(1-v^2)\,\sigma}{4v\sin^2(\dX/2)
  \sqrt{\cos^2(\dXm/2) - \cos^2((\dX)/2)}}\,,
\ee
i.e., the $\dX$ distribution has an inverse-square-root singularity at
$\dX = \dXm = 2\cos^{-1}(v) = 2.23$ for our input masses, and falls
sharply above there. This is illustrated in Fig.~\ref{fig:dchidR} where we
plot this distribution for the primary partons and for the
reconstructed jets. The low-side tail for the jets is an artifact of their
reconstruction.

\begin{figure}[!t]
 \begin{center}
\includegraphics[width=6.50in, height=3.15in]{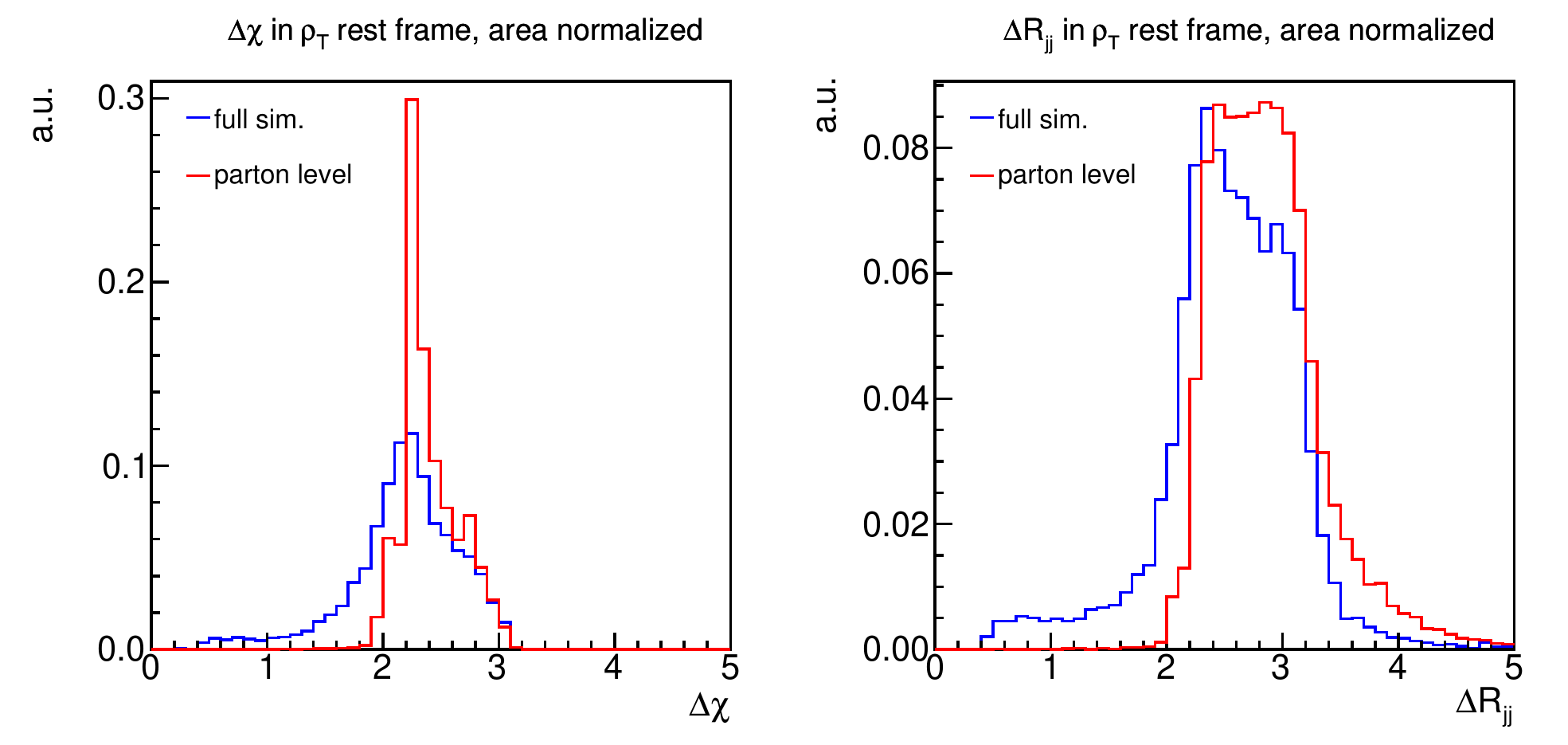}
\caption{The area-normalized $\dX$ and $\dR$ distributions for the primary
  parton/jet in $\tro \ra W\tpi$ production followed by $\tpi \ra \bar q q$
  decay, constructed as described in the text. Red: pure distribution of
  primary parton before any radiation; blue: the distribution for the jets
  reconstructed as described in Sec.~3.}
  \label{fig:dchidR}
 \end{center}
 \end{figure}

To understand this singularity better, it follows from
Eq.~(\ref{eq:cosdchi}) that $\dX$ may be expanded about $\cos\theta^* = 0$
as
\be\label{eq:dXexpand}
\dX = \dXm + \frac{a}{2}\cos^2\theta^*\ + \cdots\,,
\ee
where $a$ is a positive $v$-dependent coefficient. Then, near $\cos\theta^* =
0$, i.e., the $\dX$ threshold,
\be\label{eq:dcstardX}
\frac{d\sigma}{d(\dX)} = \frac{\sigma}{2}\, \frac{d(\cos\theta^*)}{d(\dX)}
\propto \frac{1}{\sqrt{\dX - \dXm}}\,. 
\ee
It is the simple one-variable Taylor expansion of $\dX$ in
Eq.~(\ref{eq:dXexpand}) that has caused this singularity.

The discussion of $d\sigma/d(\dR)$ for the LSTC signal shares some features
with $d\sigma/d(\dX)$, though it it is qualitatively different. The $\dR$
distribution also vanishes below a threshold, $\dRm$, which is equal to $\dXm
= 2\cos^{-1}(v)$. This remarkable feature, derived in the appendix, can be
understood simply as a consequence of the fact that the minimum of $\dR$
occurs when {\em both} jet rapidities vanish. In that case, $\dR = \dphi =
\dX$.

At threshold, however, the $\dR$ distribution is $\propto \sqrt{\dR - \dXm}$,
not the inverse square root. As illustrated in Fig.~\ref{fig:dchidR}, it
rises sharply from threshold into a broad feature before decreasing. The
measure of the $\tpi$ velocity $v$ is given by the onset of the rise, not its
peak. This is the behavior seen in the CDF data in Fig.~\ref{fig:CDFdRjj},
where the rise starts very near $2\cos^{-1}(v) = 2.23$ for our input
masses. Both the $\dX$ and $\dR$ distributions measure the $\tpi$ velocity
$v$ and, therefore, provide confirmations of the $\tro \ra W\tpi$ hypothesis
which are {\em independent} of the background under the $\MWjj$ resonant peak
and of uncertainty in the $\etmiss$ resolution as well.

The reason for this qualitative difference between the two distributions is
that $d\sigma/\dX$ involves a one-dimensional trade of $\cos \theta^*$
for $\dX$, whereas $\dR$ is parametrized in terms of the three angles
$\theta, \theta^*,\phi^*$ in an intricate way, with all three being
integrated over to account for the constraint defining $\dR$.  In contrast to
what happens in the $\dX$ case, the Jacobian singularity at the threshold is
``antidifferentiated'' twice, hence its comparatively lower strength. Using a
Fadeev-Popov-like trick, the $\dR$ distribution can be written
\be\label{eq:dsigdR}
\frac{d \sigma}{d(\dR)} =    
\int d(\cos\theta) \, d(\cos\theta^*) d(\cos\phi^*) \,
\frac{d\sigma}{d(\cos\theta^*)} \, 
\delta\left(\dR - f(\cos\theta,\cos\theta^*,\cos\phi^*) \right)\,.
\ee
The function $f(\cos\theta,\cos\theta^*,\cos\phi^*)$ is shown in the
appendix to have its absolute minimum at $\cos\theta = \cos\theta^* =
\cos\phi^* = 0$, for which its value is equal to $\dXm$. Near its minimum it
is locally parabolic and its Taylor expansion is
\be\label{eq:dRexpand}
f(\cos\theta,\cos\theta^*,\cos\phi^*) = 
\dXm + {\thalf}\left(\bth\cos^2\theta + \bthst\cos^2\theta^* +
  \bphst\cos^2\phi^*\right) + \cdots 
\ee
The positive $v$-dependent coefficients $b_{\theta}, b_{\theta^*}$ and
$b_{\phi^{ *}}$ are also given in the appendix, Eq.~(\ref{eq:bterms}). For
$\dR$ close to $\dXm$, this expansion can be used to approximate
Eq.~(\ref{eq:dsigdR}). In a similar way as for the $\dX$ distribution,
integrating first over $\cos\theta^*$ generates the appearance of a Jacobian
inverse square root singularity $\propto [2(\dR - \dXm) - (\bth\cos^2\theta +
\bphst\cos^2\phi^*)]^{-1/2}$. The two remaining integrations over
$\cos\theta$ and $\cos\phi^*$ were trivial in the $\dX$ case as the integrand
did not depend on them, but this is not so for $\Delta R$ which involves a
double integration over a restricted angular phase space defined by
\be\label{eq:omega}
 0 \leq \bth\cos^2\theta + \bphst \cos^2\phi^* \leq 2 \left(\dR - \dXm
 \right)\,.
\ee
Performing the integral in Eq.~(\ref{eq:dsigdR}) near $\dRm = \dXm$ yields a
result $\propto \sqrt{\dR - \dXm}$.

We have examined the effect of finite jet masses (as opposed to jet
reconstruction and energy resolution) on the threshold values of the $\dR$
and $\dX$ distributions and the extraction of the $\tpi$ velocity $v$ from
them. Our jets (which include $b$-jets) have masses $\simle 10\,\gev$.
Assuming, for simplicity, equal jet masses and denoting by $u =
\sqrt{1-4M_{\rm jet}^2/M_{\tpi}^2}$ the jet velocity in the $\tpi$ rest
frame, the corrected $\dXm(u)$ is
\be\label{eq:dXcorr}
\dXm(u) = \cos^{-1}\frac{v^2 - u^2(1-v^2)} {v^2 + u^2(1-v^2)} \simeq
\cos^{-1}(2v^2 - 1) -v(1-v^2)^{1/2} (1-u^2)\,.
\ee
This is less than the massless $\dXm$ by half a percent for $M_{\rm jet} =
15\,\gev$.

Finally, as noted, the $\ta$ accounts for about 25--30\% of $W\tpi$
production. This decay gives a $\tpi$ velocity of 2.00 in the $\ta$ rest
frame and $\dXm = 2.00$. The effect is clearly visible in the $\dX$ and $\dR$
distributions for the primary parton in Fig.~\ref{fig:dchidR}, but is washed
out by the low-end tails for the reconstructed jets. We believe that the low
and high-end tails are due to the two $\tpi$ jets fragmenting to three jets
and the two leading jets being closer or farther apart than the original
pair.

\section*{3. The $\tro,\ta \ra W\tpi$ mode at the LHC}

As a reminder, we assumed $M_{\tro} = 290\,\gev$, $M_{\ta} = 1.1 M_{\tro} =
320\,\gev$, $M_{\tpi} = 160\,\gev$ and $\sin\chi = 1/3$ to describe the CDF
dijet excess. The Tevatron cross section is $2.2\,\pb$. At the LHC, these
parameters give $\sigma(W\tpi) = 7.9\,\pb$ ($1.7\,\pb$ for $W\ra e\nu,
\,\mu\nu$). About 70\% of the LHC rate is due to the $\tro$; the $\tro$ and
$\ta$ interference is very small. For such close masses, it is not possible
to resolve the two resonances in the $\MWjj$ spectrum.

\begin{figure}[!t]
 \begin{center}
\includegraphics[width=3.15in, height=3.15in]{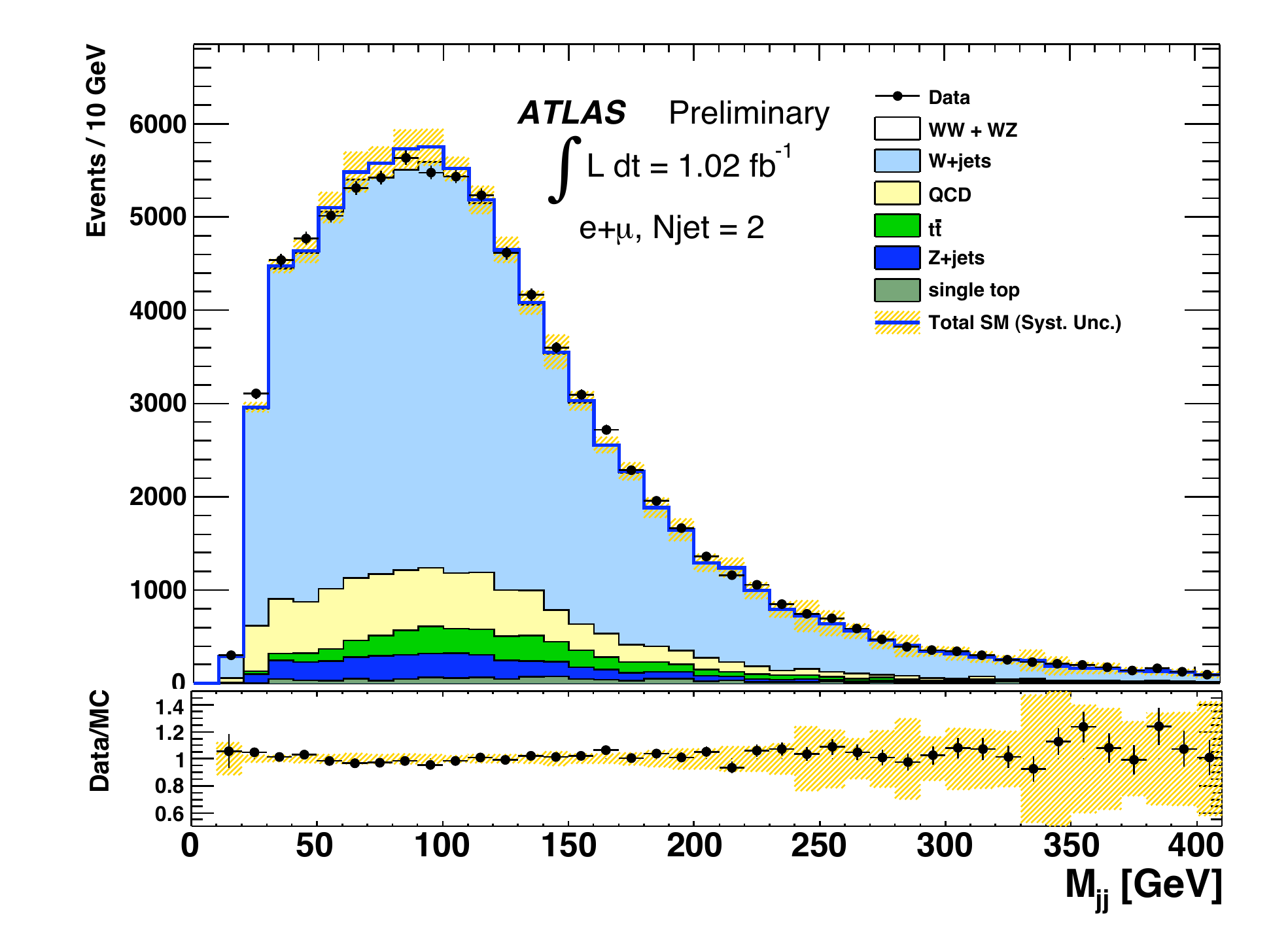}
\includegraphics[width=3.15in, height=3.15in]{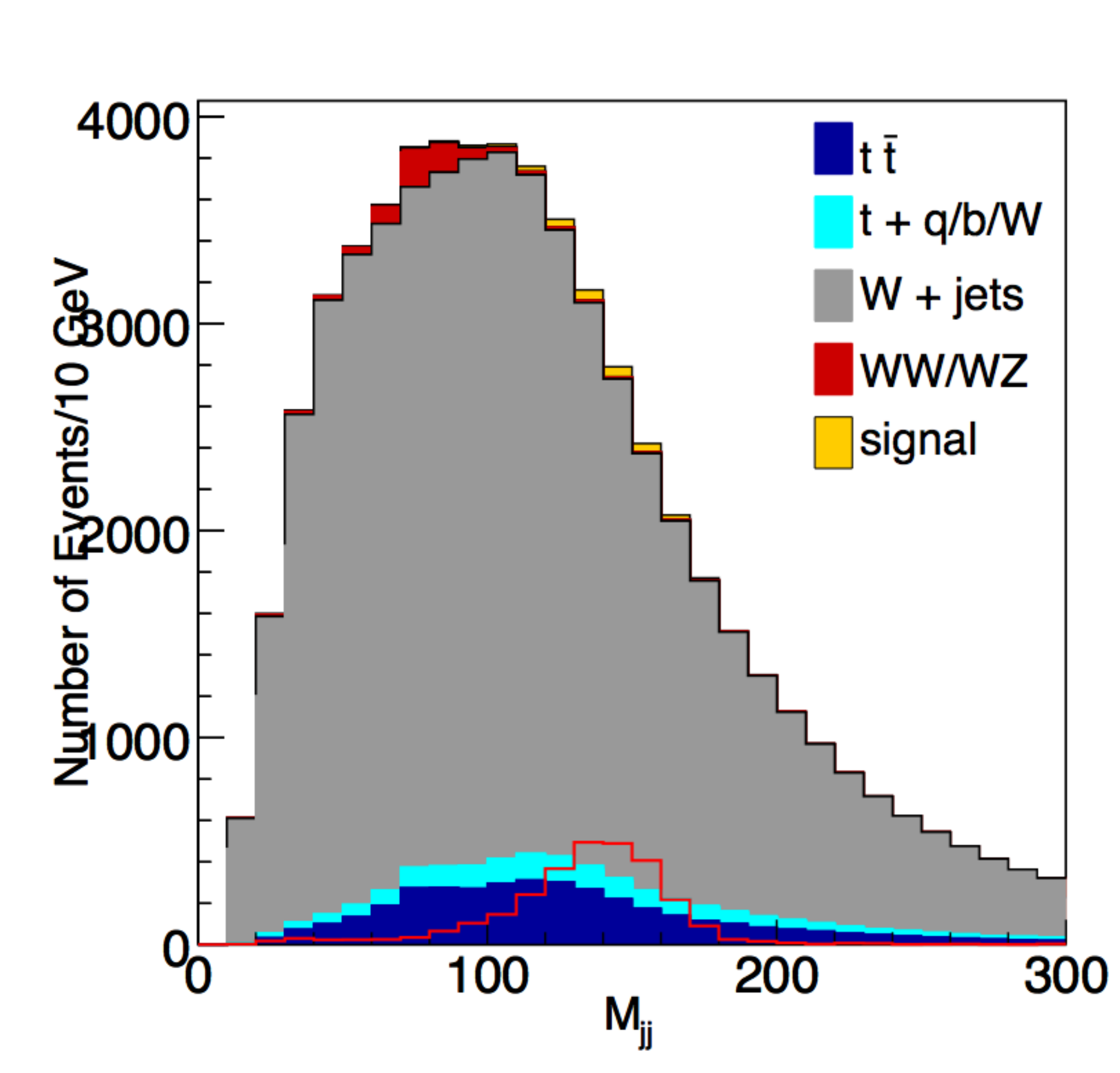}
\caption{Left: The ATLAS $\Mjj$ distribution for exactly two jets and $\int
  \CL dt = 1.02\,\ifb$; from Ref.~\cite{ATLASWjj} Right: Simulation of $Wjj$
  production at the LHC with ATLAS cuts (except that $p_T(\ell) > 30\,\gev$)
  for $1.0\,\ifb$. The open red histogram is the $\tpi\ra jj$ signal times 10.}
\label{fig:LHCWjj}
 \end{center}
 \end{figure}

 This past summer, the ATLAS Collaboration published dijet spectra for
 $1.02\,\ifb$ of $Wjj$ data with exactly two jets and with two or more jets
 passing selection criteria~\cite{ATLASWjj}. The ATLAS cuts, designed to be
 as close to CDF's as possible, were: one isolated electron with $E_T >
 25\,\gev$ or muon with $p_T > 20\,\gev$ and rapidity $|\eta_\ell| < 2.5$;
 $\etmiss > 25\,\gev$ and $M_T(W) > 40\,\gev$; two (or more) jets with $p_T >
 30\,\gev$ and $|\eta_j| < 2.8$; and $p_T(jj) > 40\,\gev$ and $\deta < 2.5$
 for the two leading jets. The $\Mjj$ distribution for the two-jet data is
 shown in Fig.~\ref{fig:LHCWjj}. There is no evidence of CDF's dijet excess
 near $150\,\gev$ nor even of the standard model $WW/WZ$ signal near
 $80\,\gev$. This is what we anticipated in Ref.~\cite{Eichten:2011xd}
 because of the great increase in $Wjj$ backgrounds at the LHC relative to
 the Tevatron. On the other hand, it is noteworthy and encouraging for future
 prospects that the ATLAS background simulation appears to fit the data well.

 In Fig.~\ref{fig:LHCWjj} we also show an updated version of our simulation
 of the LSTC $\Mjj$ signal and backgrounds at the LHC for $\int \CL dt =
 1.0\,\ifb$.  ATLAS's cuts were used except that we required $p_T(\ell) >
 30\,\gev$.\footnote{Backgrounds were generated at matrix-element level using
   ALPGENv213~\cite{Mangano:2002ea}, then passed to {\sc Pythia}v6.4 for
   showering and hadronization. We use CTEQ6L1 parton distribution functions
   and a factorization/renormalization scale of $\mu = 2 M_W$ throughout. For
   the dominant $W+$jets background we generate $W+2j$ (excl.) plus $W+3j$
   (inc.)  samples, matched using the MLM procedure~\cite{MLM} (parton level
   cuts are imposed to ensure that $W+0, 1$ jet events cannot contribute).
   After matching, the overall normalization is scaled to the NLO $W+jj$
   value, calculated with MCFMv6~\cite{Campbell:2011bn}. After passing
   through {\sc Pythia}, final state particles are combined into $(\eta,
   \phi)$ cells of size $0.1\times 0.1$, and the energy in each cell smeared
   with $\Delta E/E = 1.0/\sqrt{E/{\gev}}$. The energy of each cell rescaled
   to make it massless. Isolated photons and leptons ($e,\mu$) are removed,
   and all remaining cells with energy greater than $1\,\gev$ are clustered
   into jets using FastJet (anti-kT algorithm, $R =
   0.4$)~\cite{Cacciari:2005hq}. Finally, the quadratic ambiguity in the $W$
   reconstruction was resolved by choosing the solution with the smaller
   $p_z(\nu)$.} This tighter cut and our inability to include the data-driven
 QCD background account for our lower event rate compared to ATLAS. Despite
 this, the agreement between the two is quite good. In particular, our
 simulation shows that the CDF/ATLAS cuts cannot reveal the LSTC
 interpretation of the CDF signal at the LHC for any reasonable
 luminosity.\footnote{Models of the CDF signal that are $gg$-initiated or
   involve large coupling to heavier quarks, e.g., Ref.~\cite{Gunion:2011bx,
     Nelson:2011us}, are likely excluded by the ATLAS data.}

\begin{figure}[!t]
 \begin{center}
\includegraphics[width=3.15in, height=3.15in]{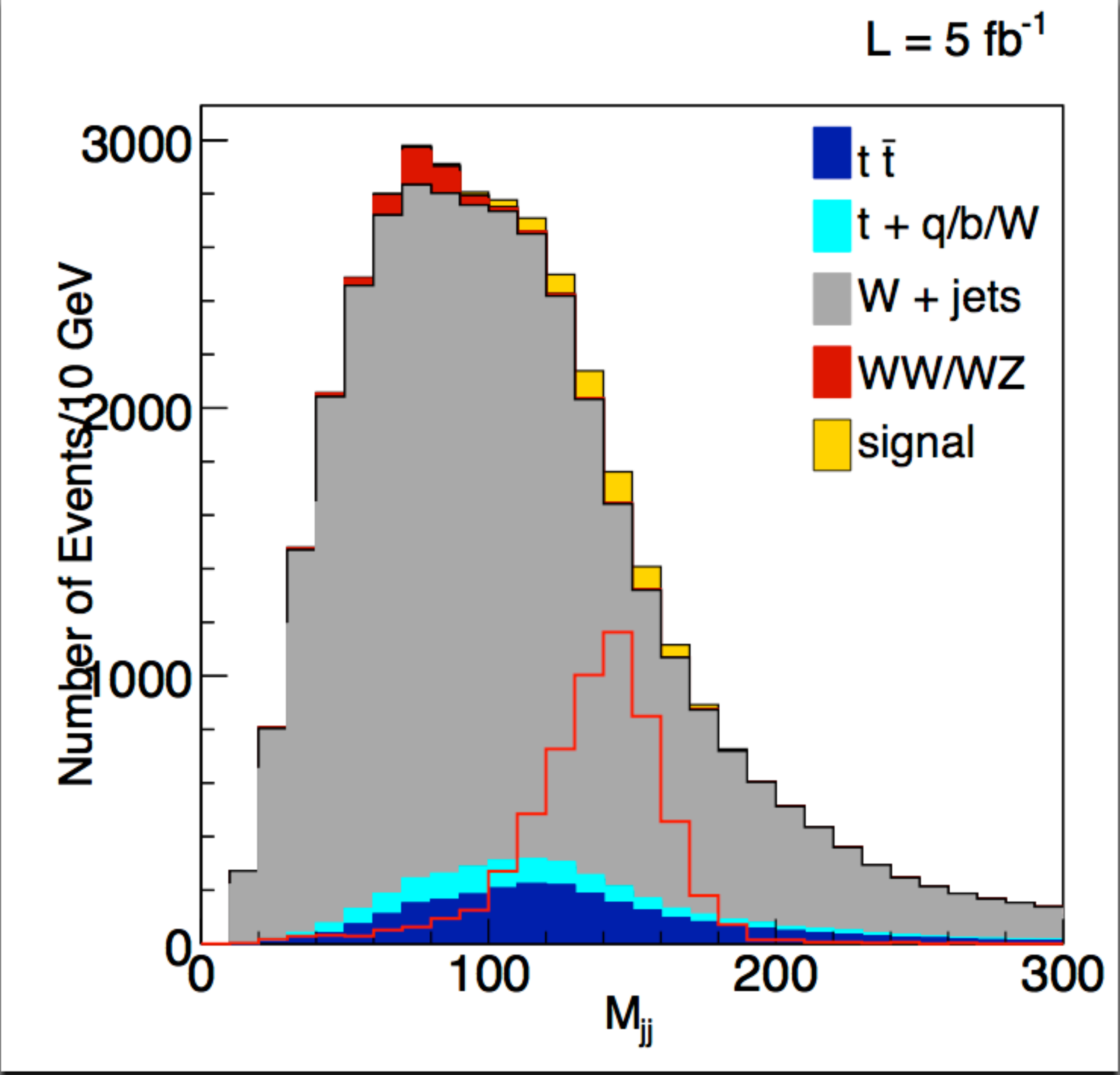}
\includegraphics[width=3.15in, height=3.15in]{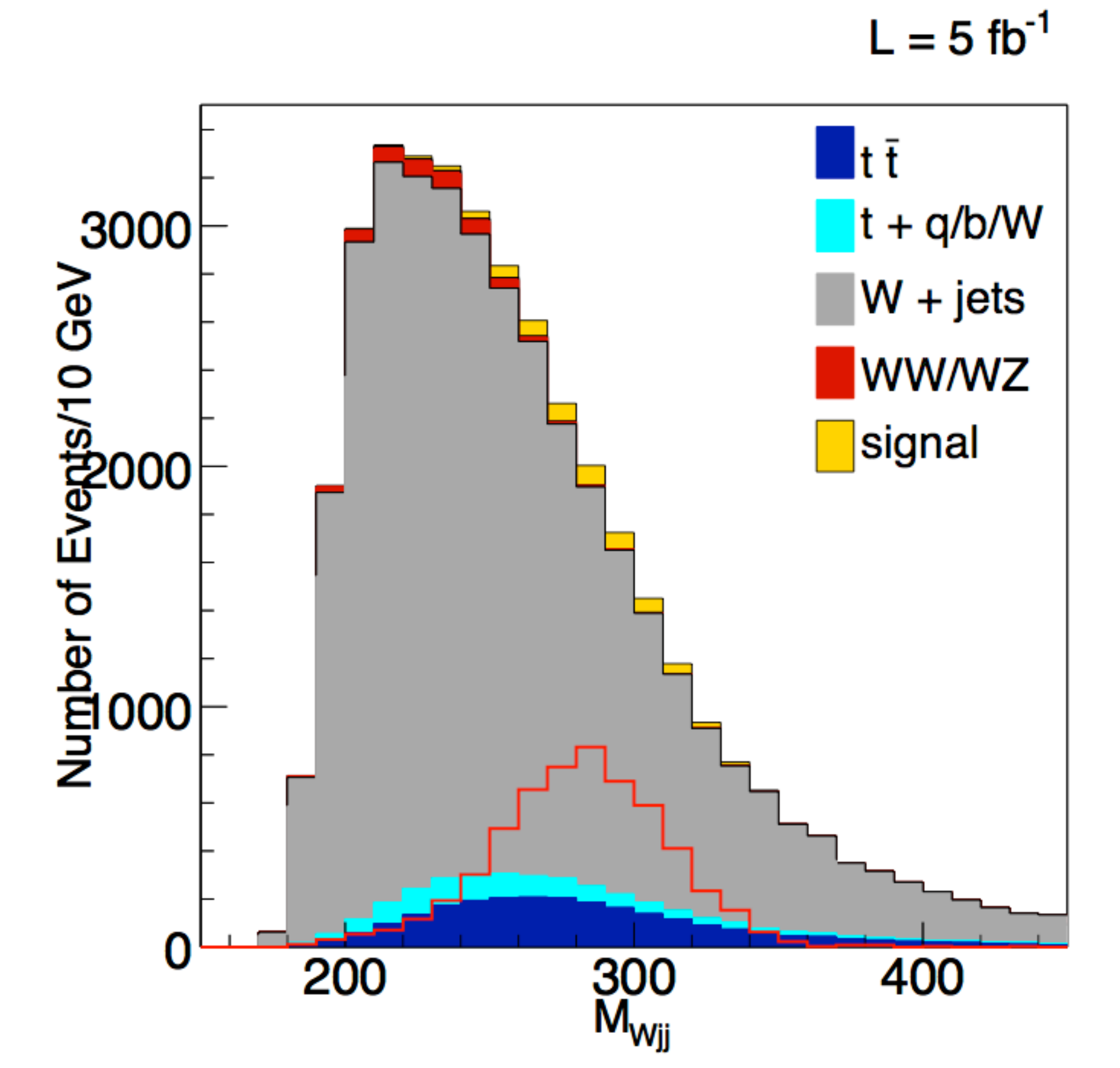}
\caption{The $\Mjj$ and $\MWjj$ distributions of $\tro,\ta \ra W\tpi \ra
  \ell\nu_\ell jj$ and backgrounds at the LHC for $\int \CL dt =
  5\,\ifb$. Augmented ATLAS-like cuts as described in the text are
  employed. The open red histograms are the $\tpi$ and $\tro$ signals times 10.
  \label{fig:ELMPWjj}}
 \end{center}
 \end{figure}

 To improve the signal-to-background, we examined a variety of cuts motivated
 by $\tro \ra W\tpi$ kinematics. Cuts quite similar to those we proposed in
 Ref.~\cite{Eichten:2011sh} typically cause the background to peak very near
 the dijet resonance. 
 To get the signal off the peak (and more like the original CDF $\Mjj$
 excess~\cite{Aaltonen:2011mk}), we used the following cuts: lepton $p_T >
 30\,\gev$ and rapidity $|\eta_\ell|< 2.5$, $\etmiss > 25\,\gev$, $M_T(W) >
 40\,\gev$ and $p_T(W) > 60\,\gev$; exactly two jets with $p_{T1} >
 40\,\gev$, $p_{T2} > 30\,\gev$, and $|\eta_j| < 2.8$; and $p_T(jj) >
 45\,\gev$, $\deta < 1.2$; and $Q = \MWjj - \Mjj - M_W < 100\,\gev$. The
 $\Mjj$ and $\MWjj$ distributions are shown in Fig.~\ref{fig:ELMPWjj} for
 $\int \CL dt = 5\,\ifb$. Counting events in the range $120 < \Mjj <
 170\,\gev$ gives $S/\sqrt{B} = 6.5$ but only $S/B = 0.050$ for this
 luminosity. Likewise, the $\dR$ and $\dX$ signals are very small and not
 useful. This is not promising. Perhaps with 20--$30\,\ifb$ a convincing
 signal could be seen in the $Wjj$ data, but it would require a very good
 understanding of the backgrounds.

\section*{4. The $\tropm,\tapm \ra Z\tpipm$ mode}

Observation of $\tropm,\tapm \ra Z\tpipm$, the isospin partner of the
$W\tpiz$ decay mode, will be an important confirmation of CDF's $Wjj$
signal. At the LHC, we predict $\sigma(\tropm,\tapm \ra Z\tpipm) = 2.5\,\pb$,
lower than $\sigma(\tropm,\tapm \ra W\tpiz) = 3.4\,\pb$ because of the
reduced phase space, $\propto p^3$. Then, $\sigma(\tropm,\tapm \ra Z\tpi \ra
\ellp\ellm jj) = 155\,\fb$ for $\ell = e$ and $\mu$. This is about 10\% of the
total $W\tpi \ra \ell\nu_\ell jj$ signal. We might, therefore, expect that
$\sim 10$ times the luminosity needed for the $W\tpi$ signal would be
required for the same sensitivity to $Z\tpi$. Actually, because there is no
QCD multijet background nor $\etmiss$ resolution to pollute the $Zjj$ data,
the situation is better than this.

\begin{figure}[!t]
 \begin{center}
\includegraphics[width=3.15in, height=3.15in]{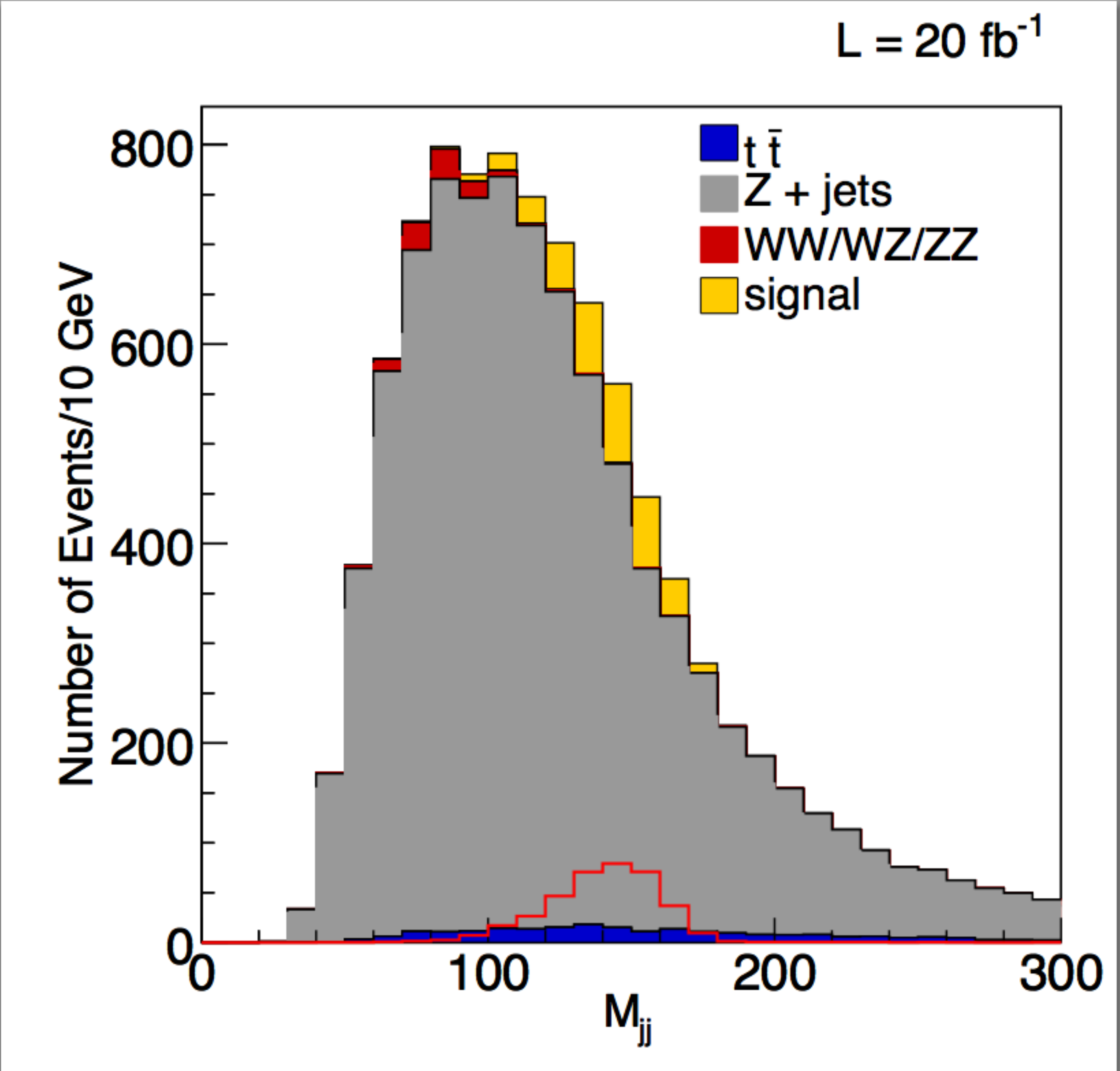}
\includegraphics[width=3.15in, height=3.15in]{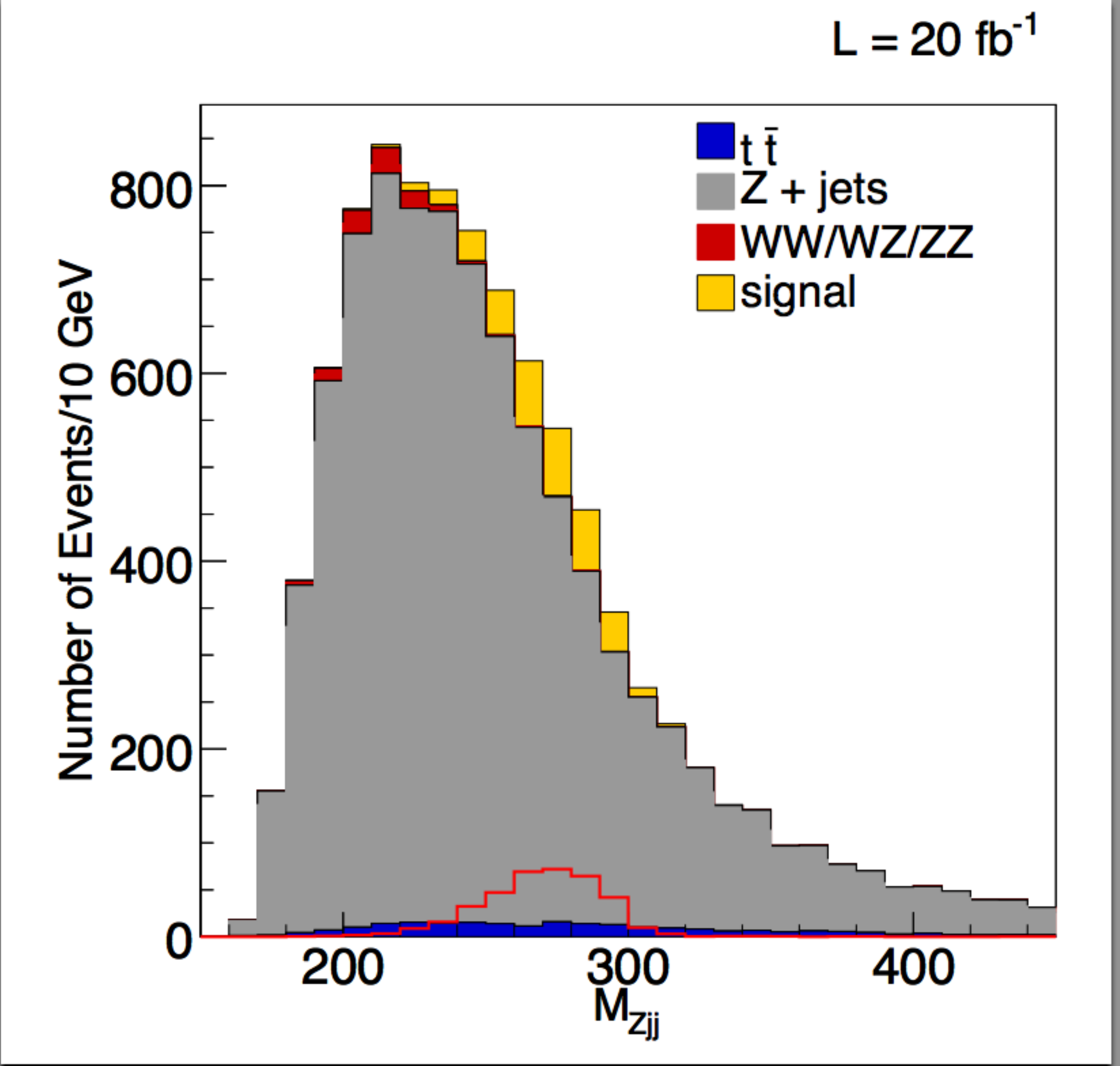}
\caption{The $\Mjj$ and $\MZjj$ distributions of $\tropm,\tapm \ra Z\tpipm \ra
  \ellp\ellm jj$ and backgrounds at the LHC for $\int \CL dt =
  20\,\ifb$. The cuts used are described in the text. The open red histograms
  are the unscaled $\tpi$ and $\tro$ signals.
  \label{fig:ELMPZjj}}
 \end{center}
 \end{figure}
\begin{figure}[!ht]
 \begin{center}
\includegraphics[width=3.15in, height=3.15in]{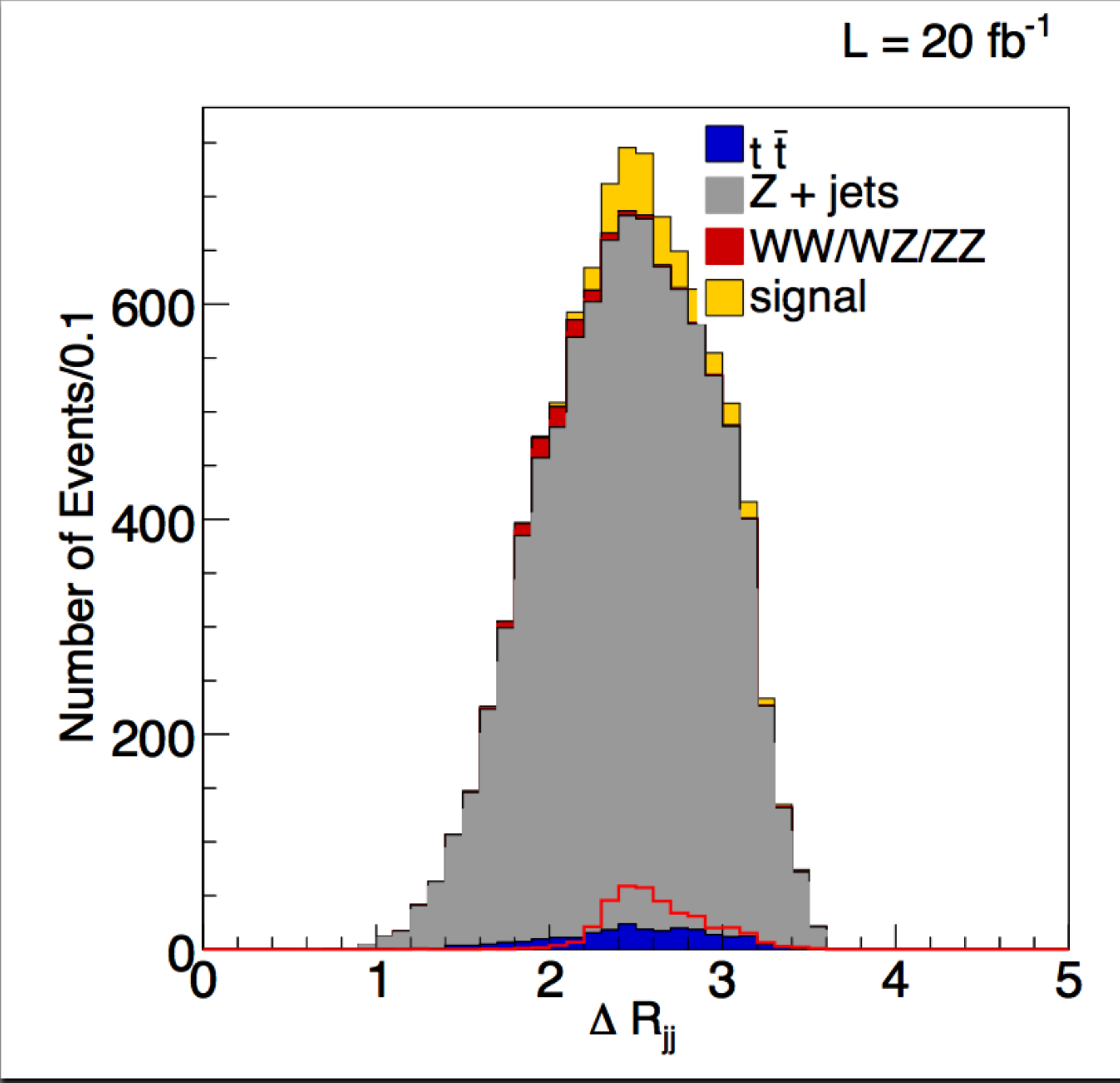}
\includegraphics[width=3.15in, height=3.15in]{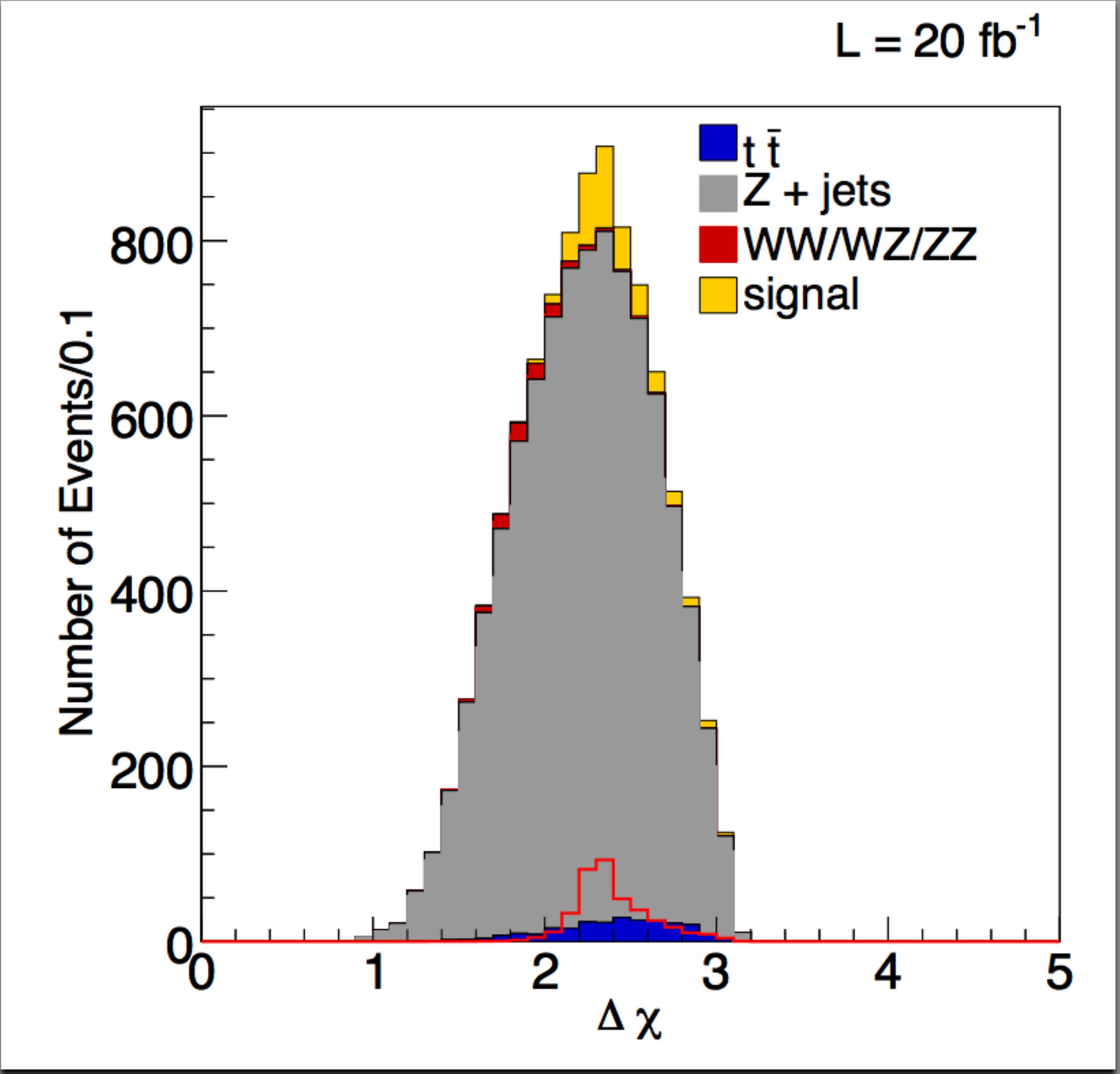}
\caption{The $\dR$ and $\dX$ distributions for $\tropm,\tapm \ra Z\tpipm \ra
  \ellp\ellm jj$ and backgrounds at the LHC for $\int \CL dt =
  20\,\ifb$. The cuts used are described in the text. The open red histograms
  are the unscaled signals.
  \label{fig:ELMPZjjRX}}
 \end{center}
 \end{figure}

 Figure~\ref{fig:ELMPZjj} shows the $Z\tpi$ signal and its background, almost
 entirely from $Z+{\jets}$, for $\int\CL dt = 20\,\ifb$. The cuts used here
 are: two electrons or muons of opposite charge with $p_T > 30\,\gev$,
 $|\eta_\ell| < 2.5$, $60 < M_{\ellp\ellm} < 100\,\gev$ and $p_T(Z) >
 50\,\gev$; exactly two jets with $p_T > 30\,\gev$ and $|\eta_j| < 2.8$;
 $p_T(jj) > 40\,\gev$, $\deta < 1.75$ and $Q = \MZjj - \Mjj - M_Z <
 60\,\gev$. This $Q$-cut is very important in reducing the
 background.\footnote{$\tro,\ta \ra WZ \ra \ellp\ellm jj$ is included in this
   simulation, but it is removed by the $Q$-cut.} These give $S/\sqrt{B} =
 6.0$ and $S/B = 0.13$ for the dijet signal in $120 < \Mjj < 170\,\gev$. The
 figure also shows the $\MZjj$ distribution; it has $S/\sqrt{B} = 5.8$ and
 $S/B = 0.11$ for $250 < \MZjj < 320\,\gev$. These signal-to-background rates
 and the position of the dijet signal on the falling backgrounds are similar
 to those in Ref.~\cite{CDFnew}. Therefore, if our interpretation of the CDF
 dijet excess is correct, both $\tpi \to jj$ and $\tro \to \ellp\ellm jj$
 should be observable in the data to be collected at the LHC in 2012.

 Figure~\ref{fig:ELMPZjjRX} shows the $\dR$ and $\dX$ distributions for
 $\tro, \ta \ra Z\tpi \ra \ellp\ellm jj$. The skyscraper-shaped $\dX$
 distribution is interesting. The background peaks at $\dX \simeq 2.3$, and
 appears rather symmetrical about this point except that its high side falls
 more rapidly above $2.7$ because $(\dX)_{\rm max} = \pi$. The signal's $\dX$
 distribution sits atop the skyscraper, concentrated in about 175~events in a
 Chrysler Building-like
 spire\footnote{\tt{http://en.wikipedia.org/wiki/Chrysler$\textunderscore$Building}}
 at $\dX = 2.2$--2.3, whereas the theoretical $\dXm = 2\cos^{-1}(v) = 2.31$
 for $\tro \to Z\tpi$. This is just as expected when the effects of jet
 reconstruction and $\ta \to Z\tpi$ are taken into account; see
 Fig.~\ref{fig:dchidR}. If the actual $\dX$ data, with our cuts, has the
 shape of our simulation, we believe the signal spire excess can be
 observed. Similar remarks apply to the shape and observability of the
 slightly broader $\dR$ distribution in Fig.~\ref{fig:ELMPZjjRX}.

\section*{5. The $\tropm,\tapm \to WZ$ mode}

Finally, the decay channel $\tropm,\tapm \to W^\pm Z$ furnishes another
important check on the LSTC hypothesis provided that $\sin\chi \simge
1/4$. The dominant contribution, $\tro \to W_L Z_L$, has an angular
distribution $\propto \sin^2\theta$ so that the production is fairly
central. We expect $\sigma(\tro,\ta \to WZ)/\sigma(\tro,\ta \to W\tpiz)
\simeq (p(Z)/p(\tpi))^3\, \tan^2\chi$. The {\sc Pythia} rates are roughly
consistent with this. For our input masses and $\sin\chi =
({\tfourth},{\tthird},{\thalf})$, we obtain
\bea\label{eq:WZrates}
\sigma(\tro,\ta \to WZ \to \ellp\ellm\ellpm\nu_\ell) &=& (13,\,21,\,44)\,\fb\,, \\
\sigma(\tro,\ta \to WZ \to \ellp\ellm jj) &=& (42,\,67,\,140)\,\fb \,,\\
\sigma(\tro,\ta \to Z\tpi \to \ellp\ellm jj) &=& (165,\,155,\,120)\,\fb \,,
\eea
for $\ell = e,\mu$.

The $\tro,\ta \to \ellp\ellm\ellpm\nu_\ell$ mode has been discussed in
Refs.~\cite{Brooijmans:2008se, Brooijmans:2010tn}. It has the advantages of
cleanliness and freedom from jet uncertainties (except $\etmiss$
resolution). Standard-model $WZ$ production at the LHC peaks at $M_{WZ} =
300\,\gev \simeq M_{\tro}$ and this is the dominant background to the
$3\ell\nu$ signal. The D\O\ collaboration searched for this channel for the
standard LSTC parameters including $\sin\chi = 1/3$, and excluded it at 95\%
C.L.~up to $M_{\tro} \simeq 400\,\gev$ so long as the $\tro \to W\tpi$
channel is closed~\cite{Abazov:2009eu}. The CMS Collaboration recently
reported a search for a sequential standard model $W'$ and for $\tro,\ta \to
WZ \to 3\ell\nu$ using $1.15\,\ifb$ of $7\,\tev$ data~\cite{CMSWZ}. The
$M_{WZ}$ spectrum and $M_{\tro}$ vs.~$M_{\tpi}$ exclusion plot are shown in
Fig.~\ref{fig:CMS}. Standard LSTC parameters, including $\sin\chi = 1/3$ were
used for this plot. Extrapolating it rules out ($M_{\tro} = 290\,\gev, \,
M_{\tpi} = 180\,\gev$) at 95\%. It appears that 5--$10\,\ifb$ will be
sufficient to exclude our CDF/LSTC mass point of $(290, 160)\,\gev$ for
$\sin\chi \simge 1/3$.
\begin{figure}[!t]
 \begin{center}
\includegraphics[width=3.15in, height=3.15in]{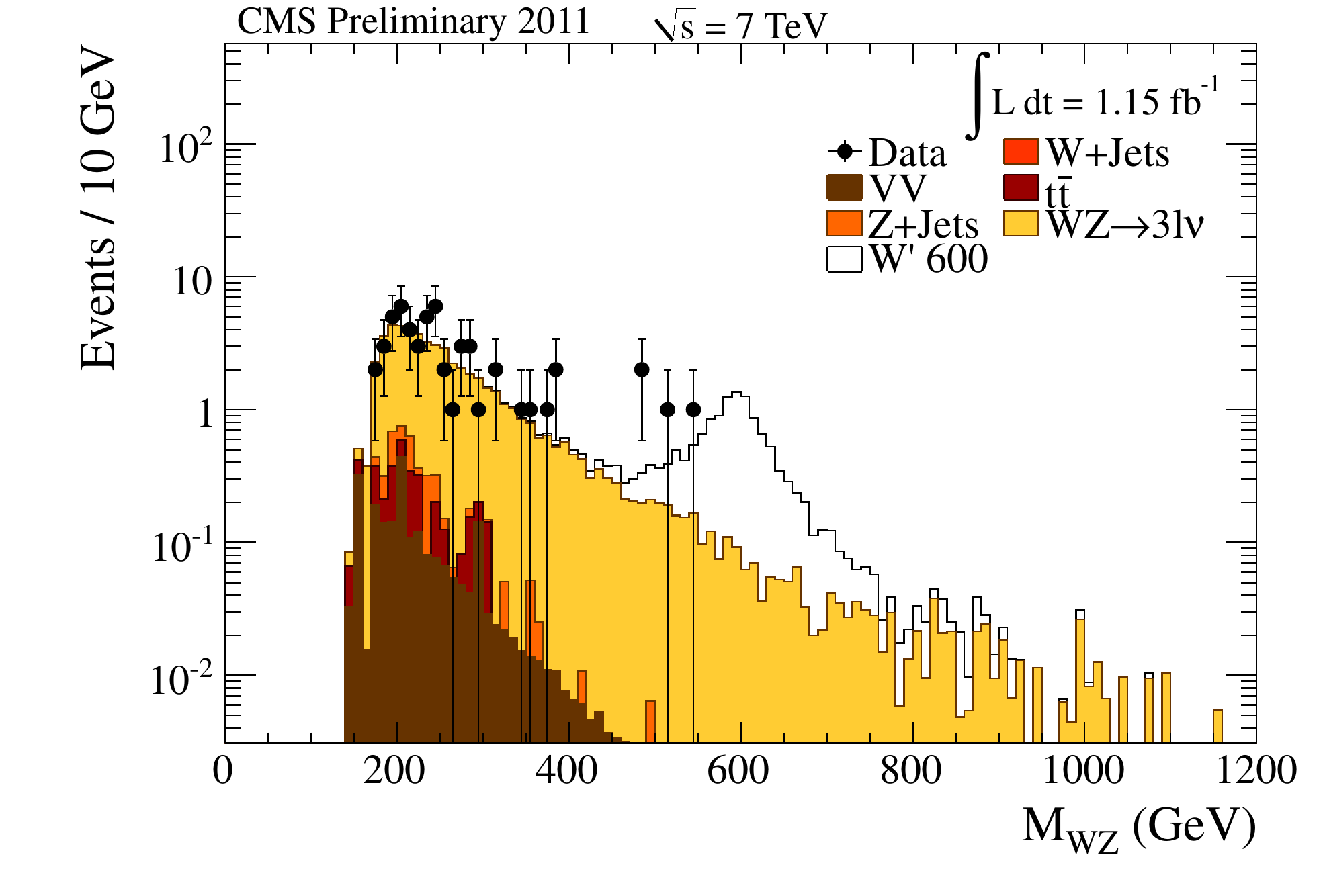}
\includegraphics[width=3.15in, height=3.15in]{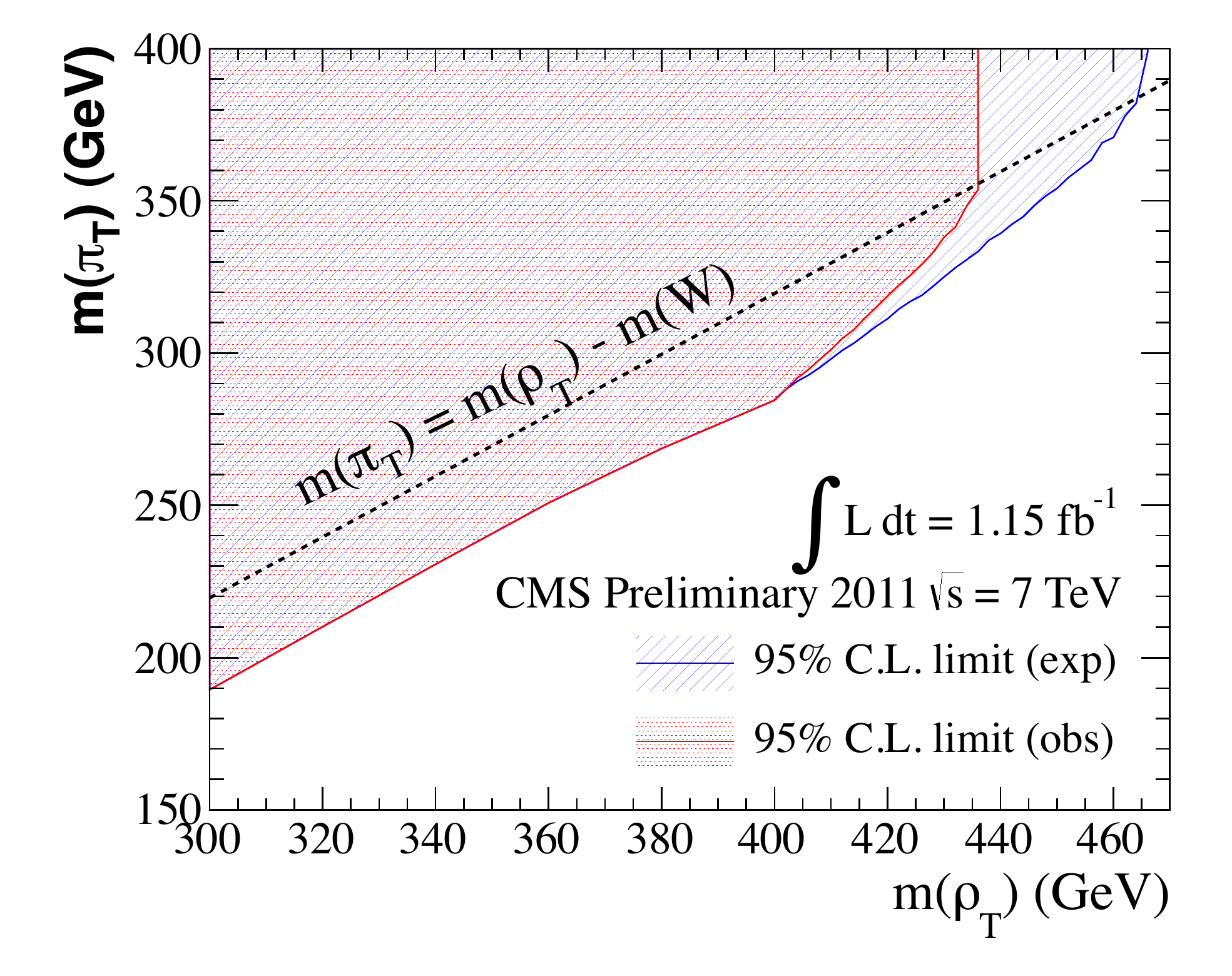}
\caption{Left: $WZ$ invariant mass distribution after the $WZ$ candidate
  selection has been optimized for the search for a $W'$ with mass of
  600~GeV. Right: Exclusion plot for LSTC as described in the text. From
  the CMS Collaboration, Ref.~\cite{CMSWZ}}
  \label{fig:CMS}
 \end{center}
 \end{figure}

 The dominant background to $\tro,\ta \to WZ \to \ellp\ellm jj$ is $Z +
 \jets$. As can be inferred from Fig.~\ref{fig:LHCWjj} for $Wjj$ production
 with ATLAS/CDF cuts, the signal will sit at the top of the $\Mjj$
 spectrum. This is what makes the dijet signal in $WW/WZ \to \ell\nu jj$ so
 difficult to see. On the plus side, since the LSTC and standard model
 diboson processes have very similar production characteristics, the two
 signals can be seen with the same cuts and will coincide. We simulated this
 mode and found what may be a promising set of cuts to extract the $W\to jj$
 signal. The basic cuts used for the $Zjj$ signal in Sec.~4 were adopted
 except that we required $p_T(Z) > 100\,\gev$, $p_T(jj) > 70\,\gev$ and $110
 < Q = \MZjj - M_W - M_Z < 150\,\gev$. The mass distributions for $\sin\chi =
 1/3$ are shown in Fig.~\ref{fig:WZMdists} for $\int\CL dt = 20\,\ifb$. The
 LSTC signal more than doubles the number of standard model $W\to jj$ events
 in the $\Mjj$ distribution and it appears that the dijet signal should be
 observable with such a data set. Including the standard diboson events gives
 $S/\sqrt{B} = 4.1$ and $S/B = 0.10$ for $60 < \Mjj < 100\,\gev$. The $\MZjj$
 signal is problematic, but it may be possible to combine its significance
 with that for $\tro,\ta \to Z\tpi \to \ellp\ellm jj$. The $\dR$ and $\dX$
 distributions are in Fig.~\ref{fig:WZRXdists}. The narrow LSTC signal and
 the diboson contribution both peak very near $\dXm = 2\cos^{-1}(v_W) = 1.21$
 and they should be observable as well.
\begin{figure}[!t]
 \begin{center}
\includegraphics[width=3.15in, height=3.15in]{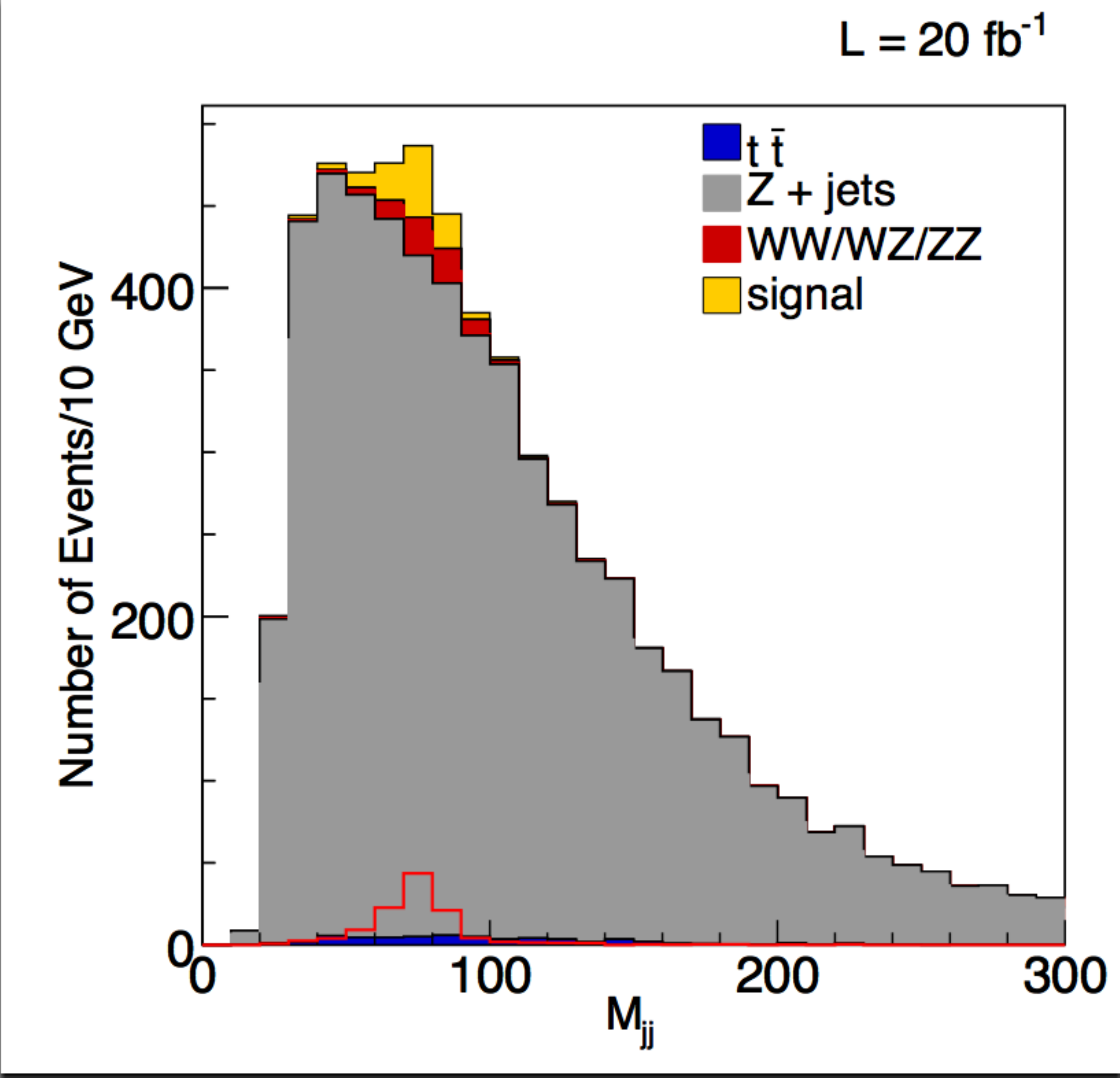}
\includegraphics[width=3.15in, height=3.15in]{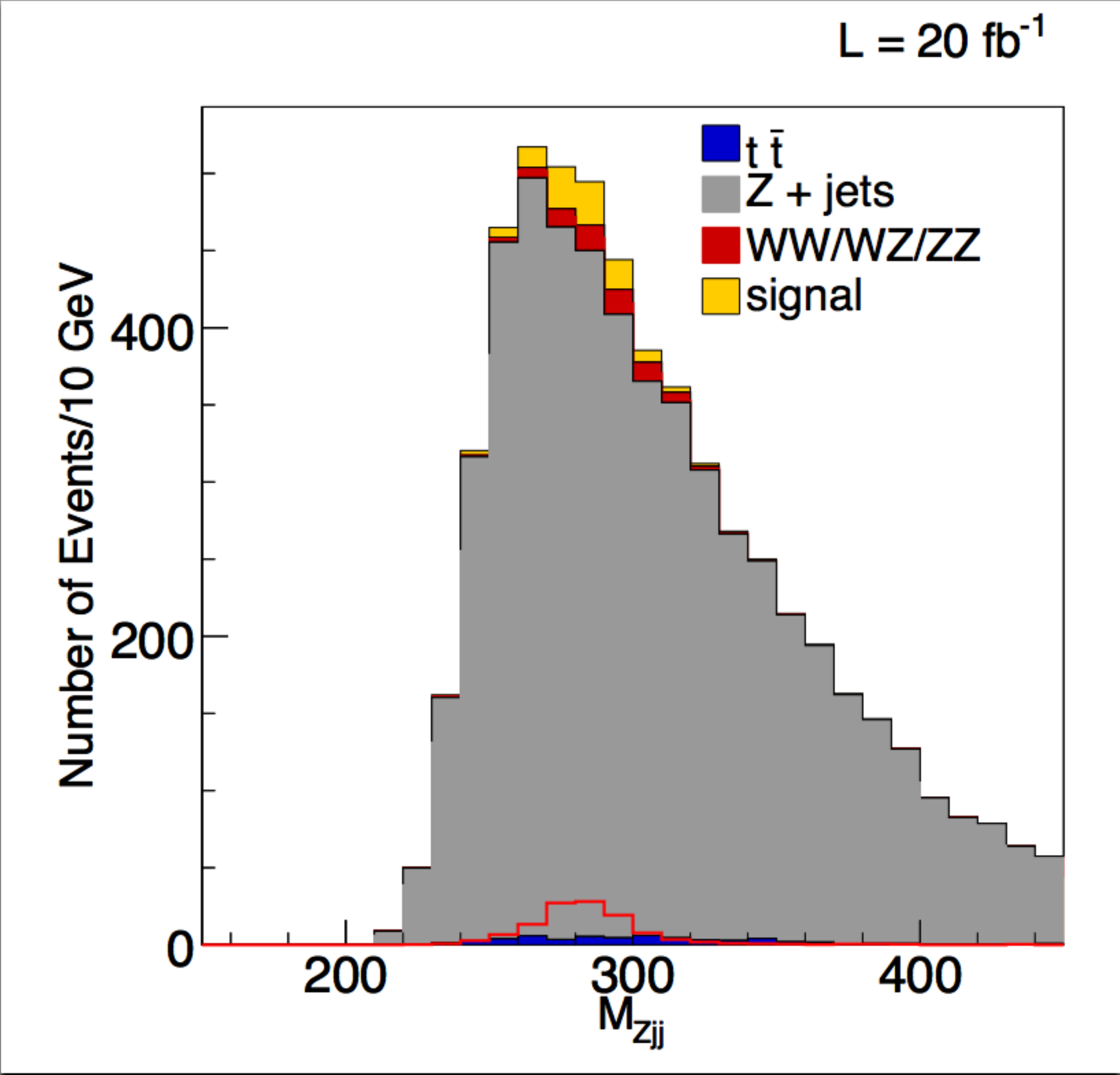}
\caption{The $\Mjj$ and $\MZjj$ distributions of $\tropm,\tapm \to WZ \to
  \ellp\ellm jj$ and backgrounds at the LHC for $\int \CL dt =
  20\,\ifb$. The cuts used are described in the text. The open red histograms
  are the unscaled $\tpi$ and $\tro$ signals.
  \label{fig:WZMdists}}
 \end{center}
 \end{figure}
\begin{figure}[!ht]
 \begin{center}
\includegraphics[width=3.15in, height=3.15in]{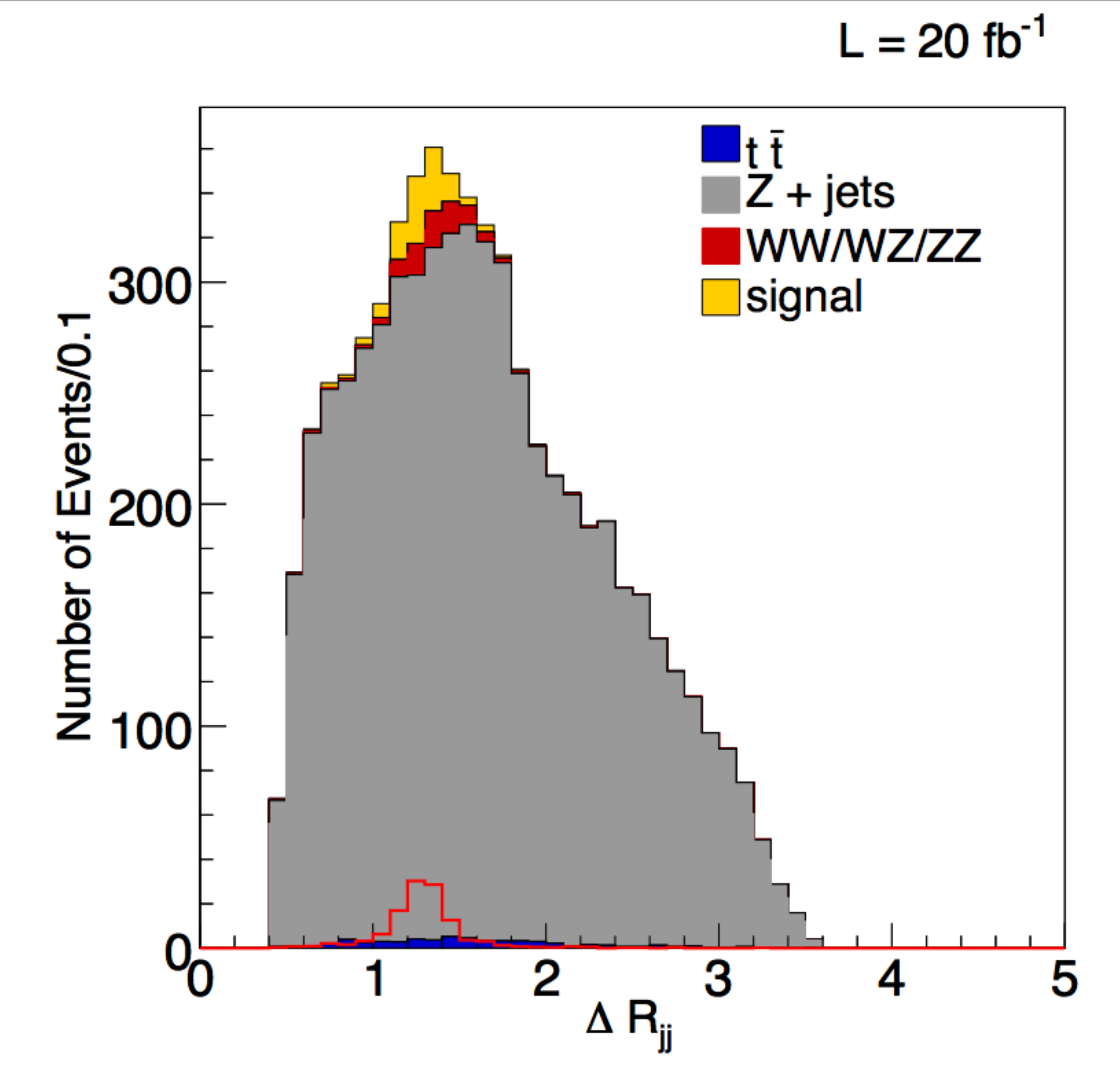}
\includegraphics[width=3.15in, height=3.15in]{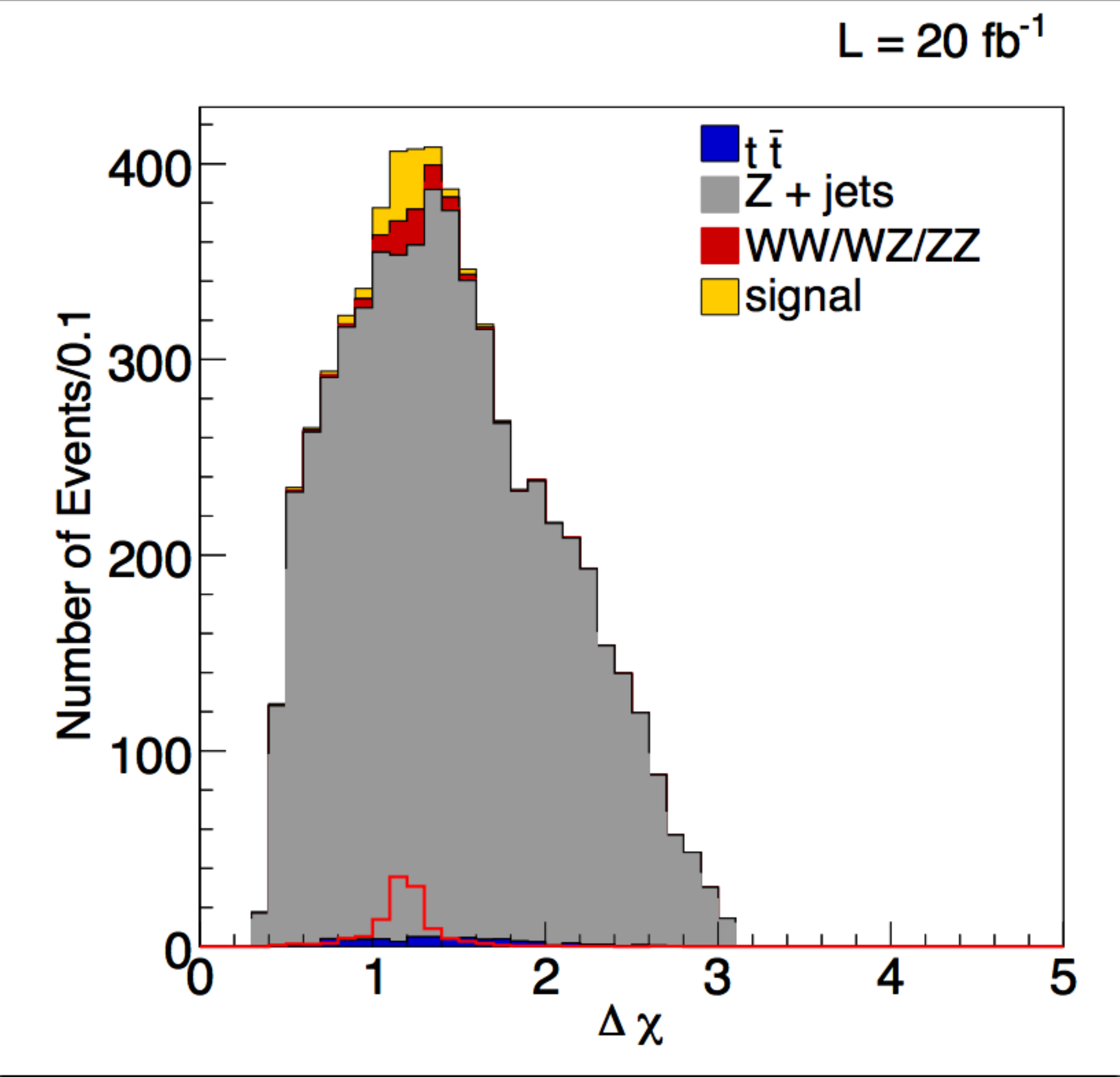}
\caption{The $\dR$ and $\dX$ distributions of $\tropm,\tapm \to WZ \to
  \ellp\ellm jj$ and backgrounds at the LHC for $\int \CL dt =
  20\,\ifb$. The cuts used are described in the text. The open red histograms
  are the unscaled signals.
  \label{fig:WZRXdists}}
 \end{center}
 \end{figure}

\section*{Acknowledgments} 

We are grateful to K.~Black, T.~Bose, P.~Catastini, V.~Cavaliere, C.~Fantasia
and M.~Mangano for valuable conversations and advice. This work was
supported by Fermilab operated by Fermi Research Alliance, LLC,
U.S.~Department of Energy Contract~DE-AC02-07CH11359 (EE and AM) and in part
by the U.S.~Department of Energy under Grant~DE-FG02-91ER40676~(KL). KL's
research was also supported in part by Laboratoire d'Annecy-le-Vieux de
Physique Theorique (LAPTh) and the CERN Theory Group and he thanks LAPTh and
CERN for their hospitality.

\vfil\eject

\section*{Appendix: Nonanalytic Threshold Behavior of $d\sigma/d(\dR)$}

\subsection*{1. Kinematics}

We recall first the definition of the angles $\theta$, $\theta^*$, $\phi^*$
and the relevant coordinate systems. Choose the $z$-axis as the direction of
the incoming quark in the subprocess c.m.~frame (or the direction of the
harder initial-state parton in the $pp$ collision).  In the $\tro$ (or $\ta$)
rest frame, $\theta$ is the polar angle of the $\tpi$ velocity ${\bs v}$, the
angle it makes with the $z$-axis.  Define the $xz$-plane as the one
containing the unit vectors $\hat{\bs z}$ and $\hat{\bs v}$, so that
$\hat{\bs v} = \hat{\bs x}\sin\theta + \hat{\bs z}\cos\theta$, and $\hat{\bs
  y} = \hat{\bs z} \times \hat{\bs x}$. Define a starred coordinate system
{\em in the $\tpi$ rest frame} by making a rotation by angle~$\theta$ about
the $y$-axis of the $\tro$ frame. This rotation takes $\hat{\bs z}$ into
$\hat{\bs z}^* = \hat{\bs v}$ and $\hat{\bs x}$ into $\hat{\bs x}^* =
\hat{\bs x}\cos\theta - \hat{\bs z}\sin\theta$. In this frame, let $\hat {\bs
  p}_1^*$ be the unit vector in the direction of the jet (parton) making the
smaller angle with the direction of $\hat{\bs v}$.  This angle is $\theta^*$;
the azimuthal angle of ${\bs p}_1^* = -{\bs p}_2^*$ is $\phi^*$:
\be\label{eq:anglestwo}
\cos\theta = \hat{\bs z}\cdot \hat{\bs v}, \quad
\cos\theta^* = \hat{\bs p}_1^*\cdot \hat{\bs v}, \quad
\tan\phi^* = p_{1y^*}^*/p_{1x^*}^*.
\ee

The jets from $\tpi$ decay are labeled $j=1,2$ and they are assumed massless.
Let $\zeta_1 = +$ and $\zeta_2 = -$, and $\cth = \cos\theta$,
$\sth = \sin\theta$, etc. The boosted jets in the lab frame are
\bea\label{eq:jetmoms}
p_j^0 &=& \thalf M_{\tpi}\gamma(1+ \zeta_j v \cthst), \nn\\
{\bs p}_{j\parallel} &=& \thalf M_{\tpi}\gamma(v+ \zeta_j \cthst)(\hat{\bs
  x}\sth + \hat{\bs z}\cth), \nn\\ 
{\bs p}_{j\perp} &=& \thalf M_{\tpi} \zeta_j((\hat{\bs
  x}\cth - \hat{\bs z}\sth)\sthst\cphst + 
  \hat{\bs y}\sthst\sphst),
\eea
where $\gamma = (1-v^2)^{-\thalf}$.

We want to find the minimum of $\dR = \sqrt{(\deta)^2 + (\dphi)^2}$ as a
function of $\cth$, $\cthst$ and $\cphst$. From Eq.~(\ref{eq:jetmoms}),
\bea\label{eq:deta}
&&\deta = {\thalf}\ln\Biggl[
  \Bigl(\frac{1+v\cthst + (v+\cthst)\cth - \gamma^{-1} \sthst\cphst\sth}
             {1+v\cthst - (v+\cthst)\cth + \gamma^{-1}
               \sthst\cphst\sth}\Bigr)\nn\\
&&\qquad\qquad  \times\Bigl(\frac{1-v\cthst - (v-\cthst)\cth - \gamma^{-1}
  \sthst\cphst\sth} {1-v\cthst + (v-\cthst)\cth + \gamma^{-1}
               \sthst\cphst\sth}\Bigr) \Biggr]\,,
\eea
and
\bea\label{eq:cdphi}
 \cos(\dphi) &=& \frac{{\bs p}_{T1}\cdot {\bs p}_{T2}}{p_{T1}\,
  p_{T2}}\\
 &=& \frac{v^2 s^{2}_{\theta} - 
       \bigl(c^{2}_{\theta^{*}}\, s^{2}_{\theta} 
     + \gamma^{-2} s^{2}_{\theta^{*}}
        \bigl(c^{2}_{\theta}\, c^{2}_{\phi^{*}} +
        s^{2}_{\phi^{*}}\bigr)\bigr)
 - 2\gamma^{-1} s_{\theta^{*}} c_{\theta^{*}} \sth\, \cth\,
 c_{\phi^{*}}}
{\Bigl\{\bigl[v^2 s^2_{\theta} + \bigl(\cthst\sth +
  \gamma^{-1}\sthst\cphst\cth\bigr)^2 +  \bigl(\gamma^{-1}\sthst\sphst\bigr)^2
  \bigr]^2 - 4 v^2\sth^2 \bigl(\cthst\sth + \gamma^{-1}\sthst\cphst\cth
 \bigr)^2\Bigr\}^{1/2}}\,. \nn
\eea
%


\subsection*{2. Minimum of $\dR$}

It clearly is hopeless to deal with the analytic expression of $\Delta R$ as
a function of $\cth$, $\cthst$, $\cphst$. However, there is a simple way to
bypass it. The quantity
\be\label{eq:Delta}
\Delta \equiv \frac{M^2_{\tpi}}{2 p_{T1}\, p_{T2}} = \cosh(\deta) - \cos(\dphi)\,,
\ee
with $\deta \ge 0$ and $0 \le \dphi \le \pi$, is a monotonically increasing
function of $\Delta R$. This is seen by parametrizing
\be\label{eq:param}
\deta = \dR \cos\lambda\,, \quad \dphi = \dR\sin\lambda\,
\ee
with $\lambda \ge 0$ and $\lambda \le \pi/2$ if $\dR \le \pi$ or $\lambda
\le \sin^{-1}(\pi/\dR)$ if $\dR > \pi$. Then
\be\label{eq:diffparam}
\frac{\partial \Delta}{\partial(\dR)} = 
\cos\lambda \sinh(\deta) + \sin\lambda \sin(\dphi) \,.
\ee
This is non-negative. It vanishes only for (1) $\dR = 0$, which means $\Delta
=0$, and this cannot happen by its definition, Eq.~(\ref{eq:Delta}), and for
(2) $\deta = 0,\, \dphi = \pi$ meaning $\Delta R = \pi$; the latter is a
saddle point. This is the ``Col du Delta'', but it is one-sided, as shown in
Fig.~\ref{fig:CdD}.
\begin{figure}[!ht]
 \begin{center}
\includegraphics[width=3.15in, height=3.15in]{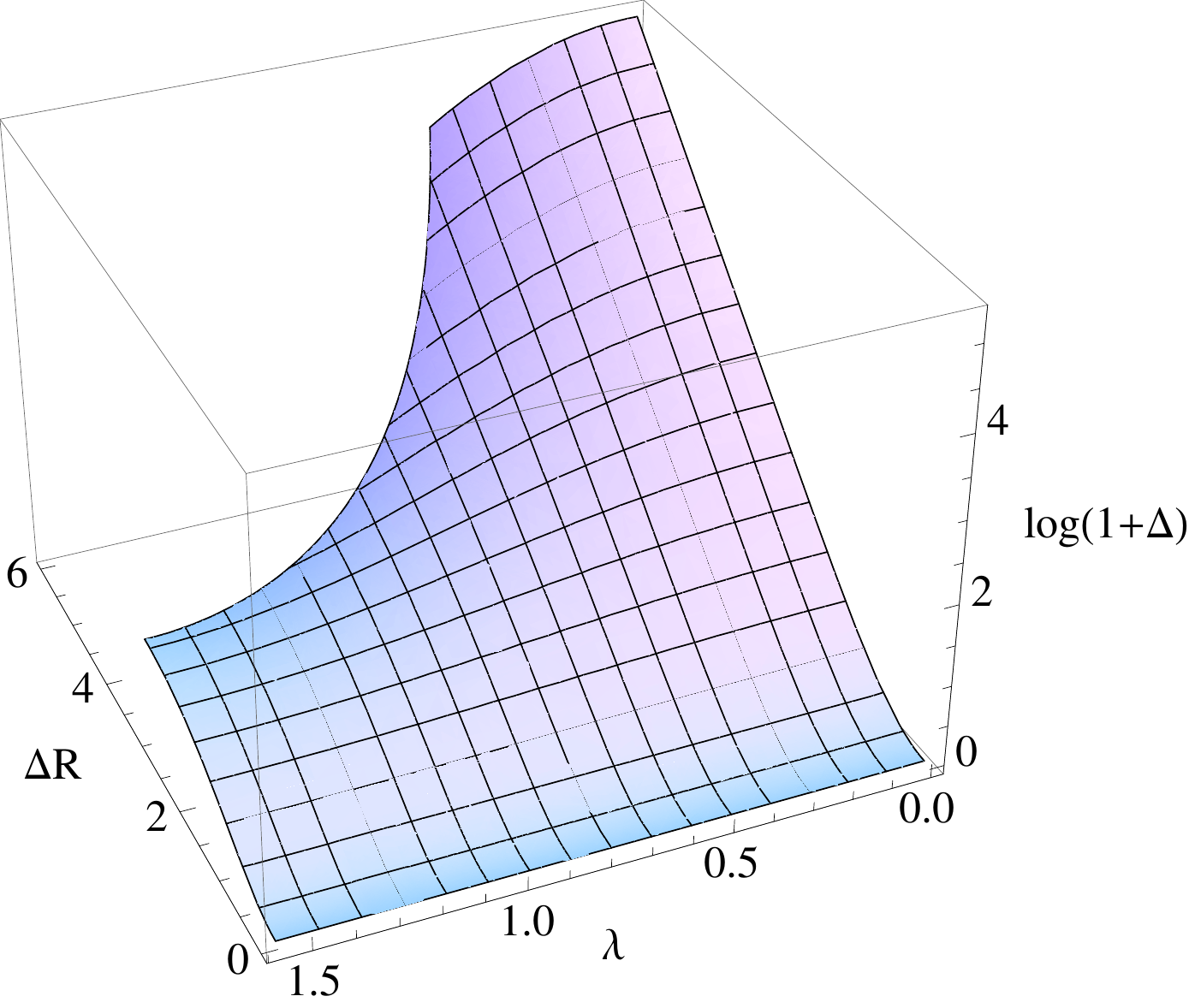}
\caption{The function $\ln(1 + \Delta)$ defined in
  Eqs.~(\ref{eq:Delta},\ref{eq:diffparam}). The Col du Delta at $\lambda =
  \pi/2$, $\dR = \pi$ is approached along the road $\lambda = \pi/2$. One
  cannot go over the pass and down the other side for the border is
  impassable. One must keep climbing along the ridge of increasing $\dR$ or
  return via the approach road.
  \label{fig:CdD}}
 \end{center}
 \end{figure}

 Minimizing $\Delta R$ thus amounts to minimizing $\Delta$, which in turn,
 amounts to maximizing $p_{T1}\,p_{T2}$. This is much simpler to examine than
 the original problem. We first maximize $p_{T1}\,p_{T2}$ at fixed $\cthst$,
 then maximize it with respect to $\cthst$. Since $p_{Tj} = \sqrt{p_{j0}^2 -
   p_{jz}^2}$ and $p_{j0}$ depends only on $\cthst$, $p_{T1}$ and $p_{T2}$
 are separately maximized at fixed $\cthst$ when $p_{1z} = p_{2z} = 0$. This
 requires $\cth = \sth\cphst = 0$. Then $p_{T1}\,p_{T2} = {(\thalf} \gamma
 M_{\tpi})^2 (1 - v^2\cos^2\theta^*)$ is maximized at $\cthst = 0$. In
 conclusion, $\dR$ is minimized if and only if
\be\label{eq:dRmin}
\cth = \cthst = \cphst = 0\,.
\ee
This corresponds to two distinct, isolated points in the angular phase space
($\phi^{*} = \pi/2, \, 3\pi/2$). The degeneracy of the minimum is only
discrete. At $\dR$'s minimum, $\deta = 0$ and $\dphi = \cos^{-1}(2v^2 - 1) =
2\cos^{-1}(v) \equiv \dXm$, so that
\be\label{eq:dRm}
\dRm = \dXm = 2\cos^{-1}(v)\,.
\ee

\subsection*{3. Local behavior around $\cos\theta= \cos\theta^* = \cos\phi^*  = 0$}

We now investigate the behavior of $\dR$ as a function of 
$\cth, \cphst$ and $\cthst$ around its minimum at $\cth = \cthst = \cphst =
0$ by means of a Taylor expansion of at most second order in any of
these variables. From, Eqs.~(\ref{eq:deta},\ref{eq:cdphi}), we obtain
\bea\label{eq:Taylor}
(\deta)^2 &=& 4 \gamma^{-2} \cphst^2 + \CO(c^3)\,,\\
\cos(\dphi) &=& \cos\dXm - (1 - \cos\dXm)\, v^2(\cth^{2} +
\cthst^2) + (1 + \cos\dXm)\, \gamma^{-2}\cphst^2 +
\CO(c^3)\,.\nn
\eea
Interpreting the latter equation as:
\be\label{eq:Taylortwo}
\cos(\dphi) = \cos\dXm - \sin\dXm\,(\dphi - \dXm) + \CO((\dphi - \dXm)^2)
\ee
we identify
\be\label{eq:dphidXm}
\dphi = \dXm + \left[v^2\tan(\dXm/2)\,(\cth^{2} + \cthst^2) - 
 \gamma^{-2}\cot(\dXm/2) c_{\phi^{*}}^{2} + \CO(c^3)\right]\,. 
\ee
Then
\be\label{eq:dRbexp}
\dR \equiv \sqrt{(\deta)^2 + (\dphi)^2} = \dXm + {\thalf}
\left(\bth\cth^2 + \bthst\cthst^2 + \bphst\cphst^2\right) +\CO(c^3)\,,
\ee
where
\bea\label{eq:bterms}
\bth &=& \bthst = 2 v^2\tan(\dXm/2) = 2v\gamma^{-1}\,,\nn\\
\bphst &=& 2\gamma^{-2}\bigl(2/\dXm - v\gamma\bigr)\,.
\eea

The shape of the surface $\Delta R = f(\cth,\cthst,\cphst)$ in the
neighborhood of the minimum $\Delta R =\Delta \chi_{min}$
is a convex paraboloid with ellipsoidal section whose eigen-directions are
parallel to the axes of the coordinates $\cth$, $\cthst$ and $\cphst$.  The
curvature is $> 0$ along each of these axes for all $0 < v < 1$; i.e. there
is no flat direction, as expected from the fact the minimum is at isolated
point(s).

\subsection*{4. Calculation of the singular part of  $d\sigma/d(\Delta R)$}

The differential cross section for $\bar q q \to \tro,\ta \to W/Z\tpi$,
followed by $\tpi \to \bar q q$ is~\footnote{Since there are two points in
  the $(\cth,\cthst,\cphst)$ phase space where $\dR$ has a minimum, $\theta
  = \theta^{*}= \pi/2$ and $\phi^{*} = \pi/2, \, 3\pi/2$, it is more
  convenient to use the variable $\cphst$ instead of $\phi^*$. This
  introduces (a) the Jacobian $(1-\cphst^2)^{-1/2}$ which is one at $\cphst =
  0$; and (b) a factor of two to account for the contributions of the two
  minima in the calculation of the normalization coefficient.}
\be\label{eq:dsig} 
d\sigma =\left[\frac{d\sigma(\bar qq \to W/Z\tpi)}{d\cth}\right] \, B(\pi_T
\to \bar qq)) 
\underbrace{
\left[\frac{1}{\Gamma(\pi_T \to \bar qq)} 
\frac{d\Gamma(\pi_T \to \bar q q)}{d\cthst\,d\cphst} \right]
}_{\left(2\pi\sqrt{1-\cphst^2}\right)^{-1}}
\,{d\cth\, d\cthst\,d\cphst} \,.
\ee
To compute the distribution in a compound variable $\zeta$, such as $\Delta
\chi$ or $\Delta R$, we use a Fadeev-Popov-like trick
\be\label{eq:FP}
1 = \int d\zeta \,\delta\left(\zeta - f\left(\cth,\cthst,\cphst\right)\right)\,.
\ee
where $f(\cth,\cthst,\cphst)$ gives the expression of $\zeta$ in terms of
the phase space variables.  The $\zeta$-distribution is then
\bea\label{eq:zetadist}
\frac{d\sigma}{d\zeta} &=&
\int d\sigma(\mbox{from Eq.~(\ref{eq:dsig})}) \,
\delta\left(\zeta - f\left(\cth,\cthst\cphst\right)\right)\,.
\eea
Let $\zeta = \Delta R$ be slightly above and close to $\dXm$, and define $\omega
= \Delta R - \dXm$ to shorten expressions. Solving Eq.~(\ref{eq:dRbexp}) with
respect to $\cthst$ gives
%
\be\label{eq:cthstsolns}
\cthst = \pm \hat{c}_{\theta^*} = \pm\sqrt{\biggl(\frac{2}{\bthst}\biggr) 
\biggl(\omega - {\thalf}\bigl(\bth\cth^2 + \bphst\cphst^2\bigr)
+\CO(c^3)\biggr)}\,.
\ee
Notice that Eq.~(\ref{eq:cthstsolns}) has to be supplemented by the
restriction
\be\label{eq:restrict}
\omega - {\thalf}(\bth\cth^2 + \bphst\cphst^2 +\CO(c^3)) \ge 0\,.
\ee
%
Substituting
\be\label{eq:subs}
\delta(\Delta R - f(\cth,\cthst,\cphst))
= (\bthst\hat c_{\theta^*})^{-1} \, \left[ 
 \delta \left( c_{\theta^{*}} - \hat{c}_{\theta^{*}}\right)
 +
 \delta \left( c_{\theta^{*}} + \hat{c}_{\theta^{*}}\right) \right] 
\Theta 
 \left[
   \omega 
   - \frac{1}{2}
   \left( 
   b_{\theta} \cth^{2} +  
   b_{\phi^{*}} c_{\phi^{*}}^{2} + o(c_{j}^{3})
  \right)
 \right]
\ee
in Eq.~(\ref{eq:FP}) and integrating over $\cthst$ leads to the following
threshold behavior for the cross section:
\bea\label{eq:dsigdRtwo}
\left(\frac{d\sigma}{d(\dR)}\right)_{\rm threshold} &\simeq&
\left[\frac{d\sigma(\bar qq \to W/Z\tpi)}{d\cth}\right]_{\cth=\cthst\cphst=0}
 \,B(\tpi \to \bar qq)\\
&&\times \frac{\sqrt{2}}{2\pi} \,
\left(\frac{1}{\bthst}\right)^{1/2}
\int d\cth d\cphst
\frac{\Theta 
 \left[
   \omega  - \frac{1}{2}\left(\bth\cth^2 + \bphst\cphst^2 + \CO(c^3)\right)
 \right]}
{\left[
   \omega  - \frac{1}{2}\left(\bth\cth^2 + \bphst\cphst^2 + \CO(c^3)\right)
 \right]^{1/2}}\,.  \nn
\eea
It is convenient to trade $\cth, \cphst$ for new variables $\rho,\kappa$:
\be\label{eq:tradevars}
\rho\cos\kappa = \sqrt{\bth/2}\,\cth\,,\quad
\rho\sin\kappa = \sqrt{\bphst/2}\,\cphst\,,\qquad
(0 \le \rho \le \sqrt{\omega}\,, \quad 0 \le \kappa < 2\pi)\,.
\ee
The integral in Eq.~(\ref{eq:dsigdRtwo}) then yields our final result, the
square-root behavior of $d\sigma/d(\dR)$ at threshold:
\be\label{eq:dsigdRfinal}
\left(\frac{d\sigma}{d(\dR)}\right)_{\rm threshold} \simeq
2^{3/2}\sqrt{\frac{\dR - \dXm}{\bth\,\bthst\,\bphst}}
\left[
\frac{d\sigma(\bar qq \to W/Z\tpi)}{d\cth}\right]_0 \,B(\tpi \to \bar qq)\,.
\ee
\vfil\eject

\bibliography{TC_at_LHC_2}
\bibliographystyle{utcaps}
\end{document}